\documentclass[preprint,11pt]{aastex}
\usepackage{psfig}

\begin{document}
\newcommand{\be}{\begin{equation}}
\newcommand{\bel}[1]{\begin{equation}\label{eq:#1}}
\newcommand{\ee}{\end{equation}}
\newcommand{\bd}{\begin{displaymath}} 
\newcommand{\ed}{\end{displaymath}}   
\newcommand{\bea}{\begin{eqnarray}}
\newcommand{\beal}[1]{\begin{eqnarray}\label{eq:#1}}
\newcommand{\eea}{\end{eqnarray}}
\newcommand{\e}[1]{\label{eq:#1}}
\newcommand{\eqref}[1]{\ref{eq:#1}}

\newcommand{\bfr}{{\bf r}}
\newcommand{\bfrp}{{\bf r'}}
\newcommand{\Scal}{{\cal S}}
\newcommand{\Jcal}{{\cal J}}
\newcommand{\Bcal}{{\cal B}}
\newcommand{\Jbar}{{\bar J}}
\newcommand{\Sbar}{{\bar S}}
\newcommand{\Jcalbar}{{\bar \Jcal}}
\newcommand{\Ibg}{{I_{\rm bg}}}

\title{Radiative Transfer and Starless Cores}
\author{Eric Keto\altaffilmark{1}, 
George B. Rybicki\altaffilmark{1}, 
Edwin A. Bergin\altaffilmark{2}, 
Rene Plume\altaffilmark{3}
}
\altaffiltext{1}{Harvard-Smithsonian Center for Astrophysics, 60 Garden Street, Cambridge
MA 02138}
\altaffiltext{2}{University of Michigan, Dept. of Astronomy, 825 Dennison Bldg,
501 East University Ave, Ann Arbor, MI, 48109}
\altaffiltext{3}{University of Calgary, Dept. of Physics \& Astronomy, 2500 University Dr. NW
Calgary, AB, T2N-1N4, Canada}

\def\etal {{et~al.}}
\def\diaz {{N$_2$H$^+$}}
\begin{abstract}

We develop a method of analyzing radio frequency spectral line
observations to derive data on the temperature, density, velocity, and
molecular abundance of the emitting gas. The method incorporates a
radiative transfer code with a new technique for handling overlapping
hyperfine emission lines within the accelerated lambda iteration
algorithm and a heuristic search algorithm based on simulated
annnealing.  We apply this method to new observations of \diaz\ in
three Lynds clouds thought to be starless cores in the first stages of
star formation and determine their density structure.  A comparison of
the gas densities derived from the molecular line emission and the
millimeter dust emission suggests that the required dust mass opacity
is about $\kappa_{1.3{\rm mm}}=0.04$ cm$^2$ g$^{-1}$, consistent with
models of dust grains that have opacities enhanced by ice mantles and
fluffy aggregrates.

\end{abstract}


\section{Introduction}

In this paper we
show how we can use 
radiative transfer modeling and heuristic search algorithms to
derive the physical conditions in molecular clouds, in particular
their temperature, density, and velocity structure,
from radio frequency spectral line observations.
We address some difficulties with this approach.
Firstly, the amount of data available from spectral line observations is
necessarily less than the amount of data required to fully describe
the clouds. Whereas the temperature, density, and velocity of the
clouds represent 15 dimensions of data, 3 dimensions each for the temperature
and density fields and 9 for the velocity vectors, the data, which
consist of the spectral line 
brightness as a function of position projected on the
plane of the sky and velocity projected on the line of sight contain
only 3 dimensions of information. 
Secondly, the parameters that we seek to find are not uniquely
related to the observed spectral line brightness. For example, 
the line brightness is related to the gas temperature and
the column density, which is itself 
a function of both the number density and the
path length or size of the cloud. Different combinations of
temperature, density, and size 
can produce the same line brightness.
Thirdly, even for simple cloud models, the
possible combinations of cloud size, density, 
temperature, and velocity structure are too many to permit an
exhaustive search.

Nevertheless we can make some progress. We can reduce the number of
parameters needed to describe the clouds by considering 
simplified geometries rather than fully
3 dimensional models. 
Within the context of these simplified models, 
some of the variables that are 
not uniquely determined by the data are much better constrained.
For example, in the context of spherically symmetric models,
both the number density and the radial size of 
a cloud can be determined from a map of the cloud.
Similarly, 
observations of multiple spectral lines, particularly hyperfine
lines within rotational transitions, can be used to separate
the density and temperature because the ratios of the hyperfine
lines are most sensitive to the column density while the ratios
of different rotational transitions are more sensitive to
the temperature. Finally, while it is still not feasible
to do an exhaustive search through the parameter space of
even simplified models,
heuristic search algorithms such as
simulated annealing are effective in narrowing down the
range of parameters and finding models consistent with the data.  
Furthermore, the history of the search through the 
available models can be used to estimate the uncertainties in
each of the parameters describing the molecular cloud.

To explore these techniques we have combined a non-LTE,
radiative transfer (RT) code  using the
accelerated lambda iteration (ALI) algorithm of Rybicki and Hummer (1991),
with a search code using an adaptation of the fast simulated annealing (FSA) 
algorithm of Szu and Hartley (1987).  The radiative transfer code generates
synthetic spectral line profiles from parameterized models of
molecular clouds while the FSA code adjusts the parameters of
the models to minimize the $\chi^2$ difference between observed
and synthetic spectra. The parameters, typically defining the
gas temperature, number density and velocity, then
represent the physical description of the cloud derived from the data.

For most molecules observed in radio astronomy, the collision rates
have been calculated only for the transitions between rotational
levels, but not for the transitions between hyperfine levels. The
hyperfine level populations therefore cannot be determined with the
full non-LTE equations.  Therefore our RT code uses an approximation
that the hyperfine levels within a single rotational level are populated
according to their statistical weights, while the rotational
transitions are treated through a more complete non-LTE calculation
(Keto 1990).  This approach also allows the use of the simpler ALI
algorithm for non-overlapping lines (Rybicki {\&} Hummer 1991), even in
cases where the hyperfine lines overlap, rather than the more complex
ALI algorithm for overlapping lines (Rybicki {\&} Hummer 1992). This
technique for treating hyperfine lines can be adopted for use with any
non-LTE radiative transfer 
code by the rather simple modification of replacing the local 
spectral line profile, usually a Gaussian, with an equivalent complex
profile consisting of the individual hyperfine lines weighted by their
relative line strengths.

We apply these techniques 
to observations of starless cores in Lynds clouds to determine
their structure. The starless cores are thought to be
gas clouds in the earliest stages of star formation that are
just beginning gravitational collapse to form protostars.
Their internal structure and the consequences of that structure
for the formation of protostars is an area of active research.
In a series of papers, Myers, Linke, \& Benson (1983), Myers \& Benson
(1983), and Benson \& Myers (1989) identified a number of dense cores in
molecular clouds as sites of future star formation.  On the basis of
IRAS observations, Beichman \etal\ (1986), separated the cores into
two classes, those containing stars and those without, and proposed
the starless cores as the earlier evolutionary phase in which star
formation had not already begun.  Based on the results of a submillimeter
continuum survey of the starless cores, Ward-Thompson \etal\ (1994)
also concluded that the starless cores were in a stage preceding star
formation. They found insufficient luminosity for accreting
protostars, and on the basis of statistical arguments found the cores
to be long-lived with timescales consistent with ambipolar diffusion
but not with gravitational free-fall as would be associated with
protostars.  However, in the case of some cores, Tafalla \etal\ (1998),
Lee \& Myers (1999), and Lee, Myers \& Tafalla (2001) came to different
conclusions.  Their observations indicated short lifetimes of
$\sim 1$ Myr and the presence of inward motions of about 0.1 kms$^{-1}$.

Two different observational surveys derived the density structure of
the starless cores.  Ward-Thompson, Motte \& Andre (1999) mapped the starless
cores in the millimeter continuum dust emission, and Bacmann \etal\
(2000) mapped the cores in 7 $\mu$m dust absorption against the infrared
background.  Both studies arrived at similar conclusions: the density
structure of the starless cores is characterized by a core-envelope
structure with an inner region of approximately constant or weakly
decreasing density and a surrounding envelope with a steep density
gradient. 

However, Evans
\etal\ (2001) and Shirley, Evans \& Rawlings (2002) argued that the
dust observations cannot be interpreted without knowledge of the
dust temperature which is not available from the observations. Therefore,
the observations may not rule out such steep density profiles after all.
They and also Zucconi, Walmsley \& Galli (2001), Stomatellos \& Whitworth
(2003), and Goncalves \etal\ (2004)
calculated that the temperature in starless cores should decrease
inward.  
Evans \etal\ (2001) suggested 
that the density profiles of Ward-Thompson \etal\ (1999)
that had been derived assuming constant dust temperature, may be
incorrect, and
the actual density profile could be much steeper,
for example as steep as $r^{-2}$.  
As for the dust absorption observations (Bacmann \etal\ 2000) that
are not affected by variations in the dust temperature and would seem
to confirm the earlier results derived from the dust emission, Evans
\etal\ (2001) argued that other errors in method invalidated these
results as well.

Different studies
came to different conclusions on the value of the 
gas density. Andre, Ward-Thompson
\& Motte (1996), Ward-Thompson, Motte \& Andre (1999), Evans \etal\ (2001),
Tafalla \etal\ (2002),
Tafalla \etal\ (2004), derived gas densities from the observations
of dust emission that were typically about an order of magnitude greater
than the densities derived from the dust absorption observations of
Bacmann \etal\ (2000).

The structure and mass of the clouds relate to 
their evolution.
Whitworth \& Ward-Thompson (2001) and Alves, Lada, \& Lada
(2001) suggest that the core-envelope
density structure may be indicative of pressure
supported Bonnor-Ebert spheres while Ward-Thompson \etal\ (1994) and
Andre, Ward-Thompson \& Motte (1996) suggest that the 
structure is consistent with
magnetic support and ambipolar diffusion. 
The flattening of the density profiles in the core centers is at odds
with models of protostellar formation that are characterized by rapid
collapse since these models predict steep density gradients approaching
the free-fall profile with $\rho \sim r^{-3/2}$ or 
$\rho \sim r^{-2}$ (Larson 1969; Larson and Starrfield 1971; Shu 1977).  
Finally, the mass of the clouds largely determines their 
stability to gravitational collapse (Bonnor 1956; Hunter 1977; 
Whitworth \& Summers 1985; Hasegawa
1988; Foster \& Chevalier 1993).

While the previous observational 
studies relied heavily on the interpretation of
dust emission and absorption to determine the density structure of the
starless cores,
molecular line observations can also yield information on the density
structure and mass as well as the velocity structure which also bears on
the evolution of the clouds.
We observed the starless clouds L1544,
L1489-NH3, L1517B in the \diaz (1--0) and (3--2) lines.  
The molecular line observations
have the advantage over the dust continuum observations in that the
temperature,
density, and velocity structures may be derived self-consistently
from the same data set. This self-consistency prevents errors
in the analysis of one independent data set from 
affecting the analyis of another data set as could be the
case if the gas-dust temperature is determined theoretically,
the gas density is determined from dust absorption observations,
and the molecular abundance based on the dust density.

To be sure,
the interpretation of molecular line observations is also
difficult, being subject to the complexities of cloud chemistry and
spectral line radiative transfer. However,  in the study of
dark clouds
the \diaz\ molecule offers some advantages that help to minimize the
uncertainties due to chemistry and radiative transfer.
Other
molecular tracers such as CO are thought to vary in abundance because
of freeze-out onto grains in the cold interiors of molecular cores 
(Brown, Charnley \& Millar 1988; Willacy \& Williams 1993; Hasegawa, Herbst
\& Leung 1992; Hasegawa \& Herbst 1993;
Caselli \etal\ 1999, 2002a, 2002b; Bergin \etal\ 2002;, Tafalla \etal\
2002).  In contrast, both observations and theory suggest that \diaz\
will maintain a relatively constant abundance with respect to H$_2$
thoughout a starless core.  (Bergin, Langer \& Goldsmith 1995; 
Aikawa \etal\ 2001;
Bergin \etal\ 2002; Caselli \etal\ 2002c).  Secondly, the hyperfine
splitting of the rotational transitions of the \diaz\ molecule caused
by the electric quadrupole interactions of the nitrogen atoms provides
additional information on the opacity of each rotational transition
that aids in constraining the excitation analysis.  With the
information provided by the hyperfine lines, it is sufficient in
principle to observe only two rotational transitions of this tracer to
separate the effects of the gas temperature and density, as well as
probe the velocity structure within the cores.

\section{Radiative Transfer and Model Clouds}
\subsection{Separating Correlated Parameters}

An advantage that the molecular line observations have over the
dust continuum is that the line observations
permit a separation of variables that are otherwise correlated. For
example, as discussed in the introduction,
the dust temperature and density can be traded off one for the other
to produce the same dust emission. Therefore, an observation of
dust emission is insufficient to define both the temperature and 
density.

Observations of two rotational transitions of a molecule that has
hyperfine structure should in principle be sufficient to determine
uniquely both the temperature and density.  As an illustration of the
principles, we consider the simple analytic equations that can be
used to derive the temperature and density under the assumption that
the cloud is a homogeneous layer in LTE.  Besides its use as an
illustration, such a calculation provides a means to derive an initial
guess necessary for the FSA search algorithm and an approximate check
on the eventual solution.  We emphasize that for the main calculations
of this paper, that derive the parameters of the starless cores, we
do not use this simple model, but use our full non-LTE radiative
transfer code.

Our illustrative model assumes the simplest 
radiative transfer through a cloud of optical depth $\tau$ with
constant source function $S$ between the upper and lower states of a
line. In this case, the intensity of a spectral line, relative to the
background intensity $\Ibg$, is
\be
  \Delta I = (I-\Ibg) = (S-\Ibg)(1-e^{-\tau})   \e{1.1.1}
\ee

This relation may be applied to two hyperfine lines, (1) and (2), of
for example, the (1--0) rotational transition of \diaz , assuming that the source
function is the same in both lines.  The optical depth in one of the
hyperfine lines, say $\tau_{\rm 1-0}^{(1)}$, can be estimated from the
observed ratio,
\be
  {{\Delta I_{\rm 1-0}^{(1)}}\over{\Delta I_{\rm 1-0}^{(2)}}}
= {{1-\exp(-\tau_{\rm 1-0}^{(1)})}\over{1-\exp(-f\tau_{\rm 1-0}^{(1)})}},
         \e{1.1.2}
\ee
where $f$ is the ratio of the absorption coefficients in the two
hyperfine lines, which is given by hyperfine theory.  
What is observed is not the line intensity itself, but rather
the antenna temperature that is 
related to the line intensity
through a multiplicative factor, $T^A = \beta\Delta I$. This factor
depends on the
characteristics of the antenna, which can usually be measured, and the
filling factor of the cloud within the antenna beam, which is usually
unknown. However, because the hyperfine lines are typically quite
close to each other in frequency, the beam sizes of  the two 
measurements will be quite close, and the multiplicative factors,
$\beta$, will cancel in the ratio of two lines,
\be
{{T_1^A}\over{T_2^A}} = {{\Delta I_1}\over{\Delta I_2}}
\ee 
Thus the ratio of the measured antenna temperatures of the
two hyperfine lines can be used to determine the optical depth.

The optical depth is related to the column density and the 
excitation temperature,\be
\tau_{1-0} = \int\alpha ds = 
{{c^2hA_{ul}N_1}\over{8\pi\nu_0\delta\nu k J(T_x)}}
\ee
where $\alpha$ is the opacity and 
$J(T_x) = (h\nu/k)/{\rm exp}(h\nu/kT_x -1)$.
The column densities of two different rotational levels are related
to the excitation temperature and the energy difference between the
levels,\be
{{N_1}\over{N_2}}={{g_1}\over{g_2}}e^{\Delta E_{12}/kT_x}
\ee
Thus from observations of two hyperfine lines of two
rotational transitions we can determine 
both the column density
and temperature.

When we consider more complex models and solve the radiative
transfer equations numerically, the same considerations that
allow the separation of the excitation temperature and density
still apply, and in principle, we can
determine the number density and temperature uniquely from
the observations. However, noise in the data, and the finite
spatial and spectral resolutions mean that there will always
be some uncertainty in the estimates of each parameter. 
There will also usually appear to be some correlation between
some parameters, but both the
uncertainties and correlations can be reduced by
better quality data.

\subsection{Searching for the best fit}

Even uncorrelated parameters are not
independent in the context of a search through multiple
parameters. For example, in fitting for the temperature
and density of a cloud, one strategy might be to run through
a series of temperatures to find the best fit, and then
through a series of densities to find the best fit density given
the recently found temperature. However, this technique will not
work since the fit of density is only good for that one particular
temperature and that temperature was only good for
the initial assumed density. Nor will 
the temperature ever budge off its previous fit unless the density is first
changed and vice versa.

An exhaustive grid search would find the
best fit, but would be too slow. 
In our study, we minimize $\chi^2$ as a
function of seven model parameters: abundance, temperature, core
density and size, the exponent of the power law for the density
in the envelope, and the infall velocity or microturbulent linewidth. 
If there are seven parameters to fit, and the range 
of each parameter is to be spanned by 10 trials, for
example sampling the temperature range of 8 to 18 K in 1 K
steps, then the
total number of trials would be $10^7$. At 100 seconds per trial,
the time required for the exhaustive grid search would be
over 10 years of computing.

Heurestic search algorithms such as simulated annealing (SA) have
been found to be effective in searching multi-dimensional
parameter spaces with reasonable efficiency and accuracy. 
In our study we use
an improved variation of SA
called fast simulated annealing (FSA), that converges on
a linear rather than a logarithmic timescale.

The SA
algorithm operates by randomly guessing new values for the parameters,
and adopting the new values if the new model has a lower $\chi^2$.  To reduce the
chance that the solution will be trapped in a local minimum of the parameter
surface and not find
the best fitting global minimum,
the algorithm is occasionally, based on a probability schedule,
allowed to adopt a new parameter value even if the computed $\chi^2$
indicates a worse fit. This
allows occasional uphill
movement and lets the search climb out of local minima. 
Convergence is controlled by reducing the
probability of acceptance of uphill moves over time. 
Thus there are two probability distributions in the algorithm,
the first is the probability of generating a new parameter 
some distance from the current value, and the second is the
probability of accepting a new model that has a $\chi^2$ higher
than the current model.
The original SA
algorithm employed a Gaussian distribution for the first probability
distribution and an exponential distribution for the second,
$\exp(-\chi^2/T)$ where $T$ is the so-called ``annealing temperature''
that is reduced over time (Metropolis \etal\ 1953). The analogy with cooling
gives the algorithm its name.  
The FSA algorithm of
Szu \& Hartley (1987) employs a Cauchy probability function in
place of the Gaussian.  The wider wings of the Cauchy distribution
allows larger jumps, searching the parameter
space faster. This in turn allows the annealing temperature to be
reduced more quickly, and thus the algorithm to converge on the
global minimum more quickly.  We further modify the original
Metropolis algorithm to allow for a search in a continuous rather than
a discrete parameter space.  This is done by setting the magnitude of
the variation of each parameter to an amount expected to produce a
change in $\chi^2$ equal to the annealing temperature. As the
annealing temperature is reduced over time, the magnitude of the
variation of the parameters is also reduced so that the searched
volume contracts in parameter space as the algorithm converges.
Because the magnitude of variation for each parameter is not known
{\it a priori} and furthermore changes as the parameters of the models
change, the variations of each parameter necessary to produce the
desired variation of $\chi^2$ are estimated from the results of the
last several iterations as the computation progresses.
A search with simulated annealing is certainly not guaranteed to find
a global minimum, but in practice the results are generally useful.

How best to estimate the uncertainty in the solution for each parameter,
or the range of each parameter that is consistent with the
data, is in practice somewhat of an open question. In theory, the range
is given by the curvature of the $\chi^2$ surface in
multi-dimensional parameter space.  However this surface is not easy
to display  once the number of dimensions exceeds 2
or 3.  For example, if we suppose that we have found the global
minimum of our parameters, then for any particular parameter, the
second derivative of $\chi^2$ with respect to that parameter, that is
the curvature, can be used as an indication of the uncertainty.  If
the $\chi^2$ surface were relatively flat along the dimension of the
parameter, then relatively large variations in the parameter would not
appreciably change $\chi^2$, the model would not be
sensitive to that parameter, and the constrained range would be
relatively large. However, even if the surface is steeply curved along
the direction of one parameter indicating a narrow range and high
precision, we cannot know from this analysis if the parameter is 
correlated with other parameters so that it would in fact have a large
range if one or more of the other parameters were allowed to
change. For example, in the case of the dark clouds, 
the abundance of \diaz\ with respect to H$_2$ and the density of
H$_2$ itself can be traded off one for the other to some
extent. However, if we have fixed the abundance, the density would
appear to be very tightly constrained indeed. If we had only 2
parameters to deal with, we could easily plot the $\chi^2$ surface,
for example as contours, to understand the correlation. However, with
additional parameters, for example the cloud size can be traded off
to some extent with the number density as well as the abundance 
to produce equivalent column
densities, and it is difficult in practice to display and understand the
correlations.

The collection of models calculated by the SA algorithm can be
organized to provide estimates of the uncertainties on each of 
the fitted parameters.
If we project the parameters of all the
multi-dimensional models onto each dimension one at a time 
the projection will
result in a 1-dimensional surface (a line) formed of the minima of the
multi-dimensional $\chi^2$ surface in projection.  The curvature of
that line describes the 
range of each parameter allowing for, or regardless
of, any variations of the other parameters that could result in a
plausible model.  The apparent ranges will be overly conservative, too
broad, because much of the variation involves correlations with parameters
that may have values far off their optima.
Nevertheless, a plot of projected $\chi^2$ 
is visually obvious, and the ranges are found
narrow enough to be interesting with respect to our understanding of
conditions in the starless cores.  

A second method of displaying uncertainties that is particularly
helpful in examining correlations among the fitted parameters is to
plot the $\chi^2$ surface as a function of 2 parameters at a time.
Correlations, for example between abundance and density, would show
up as a line of lower $\chi^2$ across the surface at an orientation
that is not parallel to either surface. Parameters that
are not correlated will have a $\chi^2$ minimum that is a single
point on the surface or a line parallel to a surface. The latter
indicates a larger range in one parameter than in the other.
Of course there may be correlations among 3 and more variables 
in the analysis, but it is difficult to display more than 2 at
a time. Examples of these plots are presented in the sections
describing the analyses of the clouds.

\subsection{ALI for molecules with hyperfine lines}

For the numerical simulation, we developed a code that employs the
accelerated $\Lambda$-iteration algorithm of Rybicki \& Hummer (1991) and
solves the radiative transfer equation in a 3-dimensional grid.  While
many such codes employ the method of short characteristics (Kunasz \&
Auer 1988) to solve the radiative transfer equation, we use a method
based on long characteristics.  In our method the rays,
distributed in angle and impact parameter, each pass through the
entire volume of the model.  The intensity field at each spatial point
is found by simple integration along each of these rays.  The various
radiation quantities required by the method are found by appropriate
frequency and angle averaging over the rays that pass through each
cell or voxel in the model cube.  Since the brightness along each ray
is calculated only twice, once in the forward and once in the reverse
direction, the computational burden of this method of long
characteristics is quite similar to that of short characteristics.
Some numerical tests of our code are described in the Appendix.

The averaging of the line radiation over frequency that produces the
local mean radiation field, $J$, requires some explanation for
molecules such as \diaz\ that show hyperfine splitting of the
rotational levels with the hyperfine lines overlapping within the
width of the line profile.  Such overlapping transitions can, in
principle, be treated with the extended formalism for the accelerated
$\Lambda$-iteration algorithm of Rybicki \& Hummer (1992), but this
requires that the collision rates between individual hyperfine states
be known.  Typically for radio frequency molecular rotational
transitions, the collision rates are known only for the total of all
the hyperfine transitions between rotational levels.  However, it
would be inaccurate to solve the radiative transfer equation ignoring
the hyperfine splitting because this would result in a single line
with an unrealistic optical depth, and in cases of moderate to high
optical depth, a saturated line with an inaccurate brightness and
width.

We use here an approximate treatment of the hyperfine transitions in
which the hyperfine levels within each rotational level are assumed to
be populated in proportion to their statistical weights.  Denoting a general
rotational level by Roman suffix $i$ and associated hyperfine sublevel by Greek
suffix $\alpha$, this assumption may be expressed,
\be
     n_{i\alpha}= {g_{i\alpha} \over g_i} n_i,    \e{2.1}
\ee
where the $n$'s are the populations and the $g$'s are the statistical
weights.  This assumption allows us to reformulate the statistical
equilibrium equations so that only the total rotational populations
$n_i$ appear.  The collisional and radiative rate coefficients between
these rotational levels will be given by averages over the hyperfine
transitions with the usual weighting by statistical weights.

In particular, the radiative rates between rotational levels $i$ and
$j$ will be properly taken into account by substituting for the
Gaussian line profile function of a single non-split line, the more
complex line profile function,
\be
    \phi_{ij}(\nu) 
  = \sum_{\alpha\beta} R_{i\alpha, j\beta} \phi_{i\alpha, j\beta}(\nu),
           \e{2.2}
\ee
that is the sum of the Gaussian profiles $\phi_{i\alpha,
j\beta}$ of each of the individual hyperfine transitions from
$i\alpha$ to $j\beta$, each weighted by the individual relative line
strength, $R_{i\alpha, j\beta}$, and each with a common linewidth that
could include both thermal and microturbulent components as well as a
velocity shift due to systematic motions.  This approximation has been
previously used in LTE radiative transfer modeling (Keto 1990) but is
also useful in non-LTE modeling.  The summed profile can then be used to
compute the radiation quantities of the ALI algorithm using the
simpler algorithm for non-overlapping lines even though the radiative
transfer equation is solved allowing for the overlapping of the
hyperfine lines.  This technique can easily be incorporated into any
ALI or Monte Carlo code by replacing the line profile function of a
single line with the line profile function representing the sum of all
the hyperfine components.

The approximation of summing the
relative intensities of the hyperfine lines
also yields a
saving in computational time for molecules such as N$_2$H$^+$ that have
overlapping 
hyperfine lines.  
This is because the computational
burden is proportional to the frequency bandwidth
and the single
profile function of the sum of overlapping lines will cover less
bandwidth than covered by the sum of the bandwidths of the individual
hyperfine lines.

The assumption of a statistical population of the hyperfine levels
might seem to be a good approximation for molecules emitting in the
millimeter radio spectrum since the energy differences of the
hyperfine levels are typically in the milli-Kelvin range whereas the
energy differences between rotational levels are several tens to
hundreds of Kelvin.  However, the \diaz(1--0) spectra of dark clouds
typically show so-called ``hyperfine anomalies'' where hyperfine lines
of the same relative strengths nevertheless are observed to have
different intensities (Caselli, Myers \& Thaddeus 1995). These
anomolies cannot be modeled by the method outlined here, and with the
lack of collision rates between hyperfine levels, it is not clear how
to improve the method.

The \diaz\ line has a particularly complex set of hyperfine components
caused by the interaction of the two nitrogen atoms in the
molecule.  It is not possible to determine by simple analytic formulas
the frequencies and relative intensities of the hyperfine components
as is the case with molecules containing a single nitrogen atom such
as NH$_3$ (Townes \& Schawlow 1955).  The hyperfine ratios and
relative strengths of all the rotational levels up to \diaz (10-9)
were calculated for us by Luca Dore of the University of Bologna using
a numerical code following Pickett (1991).  For the collision rates,
we used the published rates of Flower (1999) calculated for the
molecular ion HCO$^+$ and presumed to be similar to rates for the
molecular ion \diaz.

\section{Parameterized Models of the Starless Cores}

We describe the starless cores by simple parameterized models
based on
previous observations. 
The density structure of the starless cores is inferred from
the dust continuum observations
as a constant density core surrounded by an
envelope with a steep density gradient (Ward-Thompson \etal\ 1994;
Bacmann \etal\ 2000).  A simple model for
this density structure can be described by three parameters in
spherical symmetry: the
radius of the central region of constant density, $r_{\rm core}$, the
density in this region, $n_{\rm core}$, and 
the exponent, $\alpha$, of the power law for the density decrease in
the surrounding envelope.  While simple, this model is sufficient to
describe clouds with density profiles ranging 
from constant density to continuous power law profiles
with a singularity at the origin.

The temperature structure is suggested by previous theoretical
models indicating cooler temperatures in the center of the core
(Evans \etal\ 2001; Zucconi, Walmsley and Galli 2001;
Shirley, Evans \& Rawlings 2002; Stamatellos \& Whitworth 2003;
Goncalves \etal\ 2004).
We initially allowed our models to have two different
temperatures, a constant temperature characterizing the 
region of the constant density core, $T_{\rm core}$, and 
a second higher temperature in the envelope,
$T_{\rm env}$.
However, we found that our data were unable to
discriminate between models that had different temperatures in the core and
envelope and those that
had a single constant temperature, and we adopted the simpler constant
temperature model for all the analyses reported in this paper.  

Two cores in our study, L1489-NH3 and L1517B, have Gaussian line shapes
consistent with
a static field with a constant microturbulent broadening 
added to the thermal broadening, 
The Gaussian profiles can also be produced
by radial infall with velocities of about the same magnitude
as the broadening.
In L1489-NH3 and L1517B, velocities that are equal to the free-fall 
velocity of the enclosed
mass have about the correct magnitude, and the resulting velocity field
that decelerates inward does not produce a split line.
At the spatial resolution of the data,
the effect is simply to
broaden the line, resulting in
a spectral shape that does not deviate
significantly from a Gaussian. Thus the data for L1489-NH3 and
L1517B are consistent with
models of either turbulent support, free fall velocities, or possibly a
variety of other weak velocity fields.

One core in our study, L1544, may have an inward velocity field
(Tafalla \etal\ (1998); Williams \etal\ (1999);
and Caselli \etal\ 2002b).
For L1544, we tried models with three different velocity fields: 
1) a constant radial inward
velocity, 2) an infall velocity equal to the 
free-fall velocity $v = (GM/r)^{1/2}$ of
the enclosed mass, $M$, at each radius, $r$, and 3) 
inward velocities linearly increasing and decreasing as a function of 
radius.  
Models with the
velocity field varying as the free-fall velocity of the enclosed
mass have infall velocities that are higher in the outer parts of the
cloud  and
are therefore unable to reproduce the line splitting due to self-absorption that
is seen in the \diaz (1--0) data. Models with a velocity field in free-fall
toward a point source, such as a protostar, have inwardly accelerating
velocities that easily produce the
line splitting. However, the high velocities 
near the point source create wings on the modeled spectra
that are not seen in the data.
Models with a constant infall velocity are found to
provide better fits with some splitting and no strong wings.  It is 
possible to arrange a point source and the gas density distribution to
produce an approximately constant velocity field. A better fit is
provided by models with velocities that increase linearly from the
edge of the cloud to some maximum velocity at the cloud
center. These accelerating velocities produce
more pronounced splitting than the constant velocity case but without
the line wings from high velocity gas that is found in the free-fall
models.  In all these models, the velocity fields are 
purely diagnostic in that the velocity 
fields do not necessarily have any hydrodynamic justification, 
but any combination of density
and velocity is possible if the dynamics are time variable.

The possible variation of the abundance of \diaz\ in the starless
cores is an area of active research 
(Bergin, Langer \& Goldsmith 1995;
Aikawa \etal\ 2001;
Bergin \etal\ 2002; Caselli \etal\ 2002c; Tafalla \etal\ 2002). 
In our models
we assume that the \diaz\ 
abundance $X_{{\rm N}_2{\rm H}^+}$ 
is constant throughout the core and envelope.

The modeling reduces the description of the 
clouds to a total of seven 
parameters. In the constant density core we have the core radius
and the density. In the surrounding envelope  
we have the exponent of the density
gradient. Finally for the cloud as a whole we have the 
temperature, abundance, 
the microturbulent linewidth, and the infall velocity.
Although our radiative transfer code is three dimensional, we chose not to add
additional parameters to describe the ellipticity of the clouds because,
although none of the
observed clouds appears truely spherical in the dust continuum,
modeling the clouds as ellipses increases the complexity 
but will not improve our understanding given the
limits in our observational data. For two of our clouds we have \diaz(3--2)
data along only a single axis.

The spherical models were
generated within a volume of $0.2^3$ pc$^3$ divided into a grid of
$20^3$ or $32^3$ cells.  The spectra were computed with a frequency
spacing of 0.01 kms$^{-1}$ that was Hanning smoothed to a final
resolution of 0.02 kms$^{-1}$.  The radiation field from the model was
convolved with a Gaussian of 30$^{\prime\prime}$ to represent the
effect of the telescope beam, assuming a distance to the clouds of 150
pc.

The data are compared to the modeled spectra by computing $\chi ^2$
over both the \diaz (1--0) and \diaz (3--2) lines at four radial
positions
in L1544 that are azimuthal averages of the spectra in Figure 1, 
four positions in L1489-NH3, and three 
positions in L1517B. 

\section{Observations}

Observations of the $J =$ 3--2 transition of N$_2$H$^+$ ($\nu = 279.5117$ GHz)
were obtained using the Caltech Submillimeter Observatary 10.4m antenna
on 10 Feb.\ 1998 to 13 Feb.\ 1998.  Strip maps were completed for 3 sources
L1544 ($\alpha(2000)$ = 5:04:18.1; $\delta =$ 25:11:08),
L1489-NH3 ($\alpha(2000)$ = 4:04:47.5; $\delta =$ 26:18:42), and
L1517B ($\alpha(2000)$ = 4:55:18.8; $\delta =$ 30:38:04). 
All coordinates are J2000.
In L1544 the map positions are offset from
the continuum peak, which is located at 
$\Delta \alpha = -20''$,$\Delta \delta =-20''$
from our center
(Tafalla \etal\ 2002).   
The coordinates
of L1489-NH3 refer to the NH$_3$ peak in Benson and Myers (1989) and are
different from the coordinates of the nearby IRAS source. 
To distinguish the molecular
cloud from the IRAS source, we refer to the molecular cloud as
L1489-NH$_3$.
The millimeter dust continuum center 
in L1489-NH3 (Motte \& Andre 2001),
is found 37$''$ north of the center of our strip. 
The millimeter dust continuum peak in L1517B
(Tafalla \etal\ 2002) is offset from the center of our
strip by $-20''$,$-20''$.

The angular and spectral resolutions of these observations were
$\sim 26''$ and 0.05 km$s^{-1}$.
Pointing during the observations was monitored using Saturn with a typical
rms of $< 2''$. 
Typical
system temperatures were 250  to   350 K.
The observation of
Saturn produced a double sideband continuum antenna temperature of
T$_A^* =$ 30.5 K.  Assuming a  source brightness temperature of
135 K (Hildebrand \etal\ 1985) and an angular size of 16.8$''$ on
12 Feb.\ 1998, we
estimate a main beam efficiency of $\eta_{\rm MB} = 0.45$ for our observations.

Our observations of \diaz(1--0) in L1517B were obtained 
with the IRAM 30m antenna on 20 Aug. 1997.
This telescope has a beam size of $\sim 25''$ at 93.173 GHz,
comparable in angular resolution to the CSO beam.  The efficiency of the
IRAM observations
was measured at $B_{\rm eff} = 0.80$, and the spectral resolution was
0.13 km s$^{-1}$. 

In our analysis we combined our data with previously published
data of N$_2$H$^+$(1--0) in L1544 and L1489-NH3 from
Caselli \etal\ (2002a,b). Caselli \etal\ observed
L1544 using the IRAM
30m ($\theta_{\rm MB} \sim 25''$) and L1489-NH3 using the FCRAO
14m ($\theta_{\rm MB} \sim 56''$).  

Figures \ref{fig:bergin2} and \ref{fig:bergin1}
present the observed \diaz(3--2) spectra in L1544,
L1489-NH3, and L1517B.  
In L1544 the positions are along the major and
minor axis of the elliptical core as seen in the maps of \diaz (1--0)
emission in Williams \etal\ (1999).  In all 3 clouds the strongest
\diaz (3--2) emission coincides with the continuum
peak, and in each case we consistently see a decrease in the emission
away from the dust continuum emission peak.  In the \diaz(3--2)
spectra, the red-shoulder in the emission profile is the result of the
blended hyperfine structure of the $J$=3--2 transition rather than
self-absorption.

\section{Analysis: Individual Clouds}

\subsection{L1544}

\subsubsection{Best-fit model}

A comparison of the data and the spectra for the best fit 
accelerating infall model from a total
of several thousand trials, is shown in 
Figures \ref{fig:L1544_Model_1213} and \ref{fig:L1544_Model_1826_detail}.
The model parameters are summarized in Table 1.
Figures \ref{fig:L1544_Model_1826_fine_detail} 
and \ref{fig:L1544_Model_1213_fine_detail} show the inner 3 hyperfine lines
of the (1--0) transition for the best fit linearly accelerating infall model
and for comparison, a best fit spectrum from a 
constant velocity model. The comparison suggests that the gas in
L1544 is not only infalling but also accelerating inward. The absence of
line wings on the observed spectra
requires that the maximum infall speed is less than 0.2 kms$^{-1}$.

Figure \ref{fig:chi2_l1544} shows the $\chi^2$ values of all the
models calculated in the search projected onto each of the fitted
parameters.  
As described earlier, 
the curvature of an imaginary line through the
lowest points suggests the range of acceptable values for each
parameter. And at each value of a parameter, all the points in 
a vertical line at the value represent the $\chi^2$ of all the
models calculated with that value. 
For example, there are hundreds of different models calculated with a 
temperature of 10K but with many different variations of all the other parameters.
The lowest $\chi^2$ at 10K represents the best model achieved with
that temperature allowing for any and all variations of the other parameters. 

Some of the parameters are more tightly
constrained than others. 
All models with densities below $2 \times 10^5$ cm $^{-3}$ and above 
$10^6$ cm $^{-3}$ result in significantly worse fits to the data. 
The range of acceptable abundances 
is between 
$6 \times 10^{-10}$ to $3 \times 10^{-9}$.   
The gas temperature 
cannot be much colder than 7 K, since models with
low temperatures do not provide good matches to the spectra
even if the density is increased in the core.    
The $\chi^2$ surface projected onto the axis of the core radius
is sparsely populated below 0.002 pc both because core radii below 0.002 pc do not fit
the data well and because the volume of the phase space below 0.002 pc is small
compared to the volume above 0.002 pc.  
(Phase space volumes can be adjusted mathematically.
For example, we fit the logarithm of the density and abundance to allow a relatively
greater volume in phase space at lower parameter
values.)
The density power law 
in the envelope must be steeper than $r^{-1.2}$ but 
the significantly worse fits for core radii below 0.002 pc 
imply that the density cannot
increase as a power law to very small radii.  

Possible correlations among the parameters are explored by plotting
the surface of $\chi^2$ as a function of two parameters at a time.
Figure \ref{fig:chi2_surface_l1544} shows that
there are no strong correlations among the fitted parameters. 

\subsubsection{Comparison with other results}

Both Ward-Thompson \etal\ (1999)
and Bacmann \etal\ (2000) derived 
a density structure from dust observations of 
L1544 consisting of a constant density core
and variable density envelope, although
the latter proposed a steeper
density gradient in the envelope. This difference appears 
reconcilable.
The Bacmann \etal\ (2000) results were derived from
observations that cut across
the short axis of the elliptical cloud whereas the Ward-Thompson
\etal\ (1999) results were derived from annular averages of a
grid of data.

The core density from our best fit model, 
$n_{\rm H_2} = 5   \times 10^5$ cm$^{-3}$, is below 
that derived from mm and sub-mm dust emission observations,
$10^6$ cm$^{-3}$ (Ward-Thompson, Motte, \& Andre 1999; Evans \etal\ 2001;   
Tafalla \etal\ 2002), but comparable to the value derived from dust
absorption
$4 \times 10^{5}$ cm$^{-3}$ (Bacmann \etal\ 2000).
The gas densities derived from the dust emission observations assumed 
a dust mass opacity of $\kappa_{1.3{\rm mm}} = 0.005$ cm$^{-2}$ g$^{-1}$.
If one assumes a higher opacity of 
$\kappa_{1.3{\rm mm}} = 0.04$ cm$^{-2}$ g$^{-1}$ a factor of 2 above
the millimeter opacity
suggested by Kruegel and Siebenmorgen (1994) but still a factor of 25 
below that 
proposed by Ossenkopf and Henning (1994), then the densities derived
from the dust emission observations would be in agreement with the densities
derived from both the dust absorption and the molecular line observations.

Our derived abundance, $X$(\diaz) $\sim 10^{-9}$, 
is in agreement with Caselli \etal\ (2002b), but
higher than proposed by Tafalla \etal\ (2002),
$X$(N$_2$H$^+$) $\sim 8 \times 10^{-11}$.  The result of 
Tafalla \etal\ (2002) is based on their H$_2$ densities derived from
dust emission, and their 
lower abundance reflects the higher density given by the dust.
In contrast, we use our observations of the \diaz(3--2) line and combine these
with the same \diaz(1--0) observations of Tafalla \etal\ (2002) to derive both
the density and abundance directly from the molecular data alone.
Our $J$= 3--2 data do not 
allow the high densities as found from mm continuum
emission and to match the intensity of $J$= 1--0 emission requires a higher
abundance.

Our core radius (called R$_{\rm flat}$ 
in Ward-Thompson \etal\ 1999; Bacmann \etal\ 2000) 
in our best fit model is 1000 AU, and, considering our angular 
resolution (30" = 4500 au at 150 pc), comparable to the 1900 AU
deduced by Bacmann \etal\ (2000).  

Our best fit density
power law is consistent
with the envelope power laws derived from the dust observations
of Ward-Thompson \etal\ (1999) and Bacmann \etal\ (2000).
Evans \etal\ (2001) suggested that the dust continuum emission, 
when modeled
with a temperature structure decreasing from edge to center, might also be fit 
by a density 
power law consistent with the singular isothermal sphere.  We find that our data
do not support this hypothesis.   
As a further check, we also ran some models 
with
the temperature fixed at  5 K within the core
radius, and allowing a different temperature, as a fitted parameter, 
within the power-law envelope. 
We also searched models with
the abundance  described by a power-law
that decreases with radius ($X = X_0 r^{-m}$).  In both cases the fits
to the data did not improve, and the estimated densities were not appreciably
higher.   Thus in our analysis we find good agreement 
with the suggestion that the 
density profile in the pre-stellar cores flattens towards the center. 

One way to understand the constaint against high densities
even in the models with lower temperature 
is to suppose that the emission were 
optically thin, the populations in
LTE, and the Rayleigh-Jeans approximation applicable. Then to first
order, the line brightness would be equal to the product of 
the optical depth and the 
source function which is itself
proportional to temperature. In this case 
a change in the temperature by a factor of 2 or 3 
would only allow a proportional change in the density 
if  the  line brightness were to
remain the same.
This would be insignificant in terms of a steep ($r^{-2}$) power law.

\subsection{L1489-NH3}

\subsubsection{Best-fit model}

The model for L1489-NH3 is similar to L1544.
However, because the  observed
spectra in L1489-NH3 are consistent with Gaussian profiles and
show no evidence for emission self-reversals indicative of infall
as seen in L1544,
we used a model for L1489-NH3
with a static velocity field, $v(r) \equiv 0$, and
included a
microturbulent width as a parameter.   
The spectra of the best fit model are shown in Figures
\ref{fig:L1489_Model_955} and \ref{fig:L1489_Model_955_detail}, 
the parameter values are given in Table 2,
and the projected $\chi^2$
plots are given in 
Figures \ref{fig:chi2_l1489} and  \ref{fig:chi2_surface_l1489}.

The lower values of $\chi^2$ in the 
L1489-NH3 models compared to L1544 is a
result of the lower signal-to-noise ratios
of the L1489-NH3 spectra compared to L1544.  
The reason why
variations of
$\chi^2$ of less than unity are still significant can be
understood by examination
of the least squares fitting process.
In the modeled and observed spectra, signal  
is expected only in certain channels at specific frequencies 
depending upon the hyperfine structure, velocity field, and line width.   
Channels with no signal may be considered significant or not, depending
on where emission is expected. However, it is not possible to know
or specify with certainty which channels in the model and data 
are significant in order to include only the significant 
channels 
in the computation of $\chi^2$. Channels that have no significant
line emission but only noise will contribute an average  value of
unity to the $\chi^2$. 
In spectra which are
noisy (such as L1489-NH3 and L1517B) the relatively larger number of
channels that are free of emission above the noise
produces values of $\chi^2$ that
are all closer to unity and with a smaller
variation in  $\chi^2$ between the models. Nonetheless, the 
relative variations
in $\chi^2$ are still significant, and for example, models with 
$\chi^2$ of 1.5 are noticeably
much worse than the best fit model with a
$\chi^2$ of 1.2.

Figure \ref{fig:chi2_surface_l1489} 
explores the possible correlations among the parameters.
Here because of the lower signal to noise data, and the lower
ratio of the cloud size to angular resolution, our analysis is not
as successful in uniquely characterizing the individual parameters.
The plots show correlations between the abundance and each of the
three parameters, the temperature, density, and core size. Still
the ranges of the fitted parameters, as seen in either the
plots of the projected $\chi^2$ surface or the contoured surfaces
are narrow enough to be interesting relative to our current knowledge 
of these clouds. We expect that better data would result in more
tightly constrained parameters.

\subsubsection{Comparison with other results}

For L1489-NH3 there is no previous
analysis with regards to the density structure from dust emission
or absorption in the literature.   The cloud was mapped in dust emission
by Motte \& Andre (2001), who estimated a total flux of 55 mJy/beam towards
the peak.  Using this flux we estimate the average density within the
13$''$ beam of $10^{6}$ cm$^{-3}$ 
(assuming $\kappa_{1.3{\rm mm}} = 0.005$ cm$^2$ g$^{-1}$ and a dust
temperature of 10 K).
Assuming that this average value is a lower limit to the density in the core
of a core-envelope structure, we find that,
similar to L1544, the density, $n_{core}=10^5$ cm$^{-3}$, derived from
N$_2$H$^{+}$ is below that estimated via dust emission. 
Using the higher dust mass opacity of 
$\kappa_{1.3{\rm mm}} = 0.04$ cm$^2$ g$^{-1}$ the gas densities estimated
from
the dust and molcular line observations would be consistent.
Our estimated temperature of 17 K is higher than, but, as indicated
by the $\chi^2$ distribution in Figure \ref{fig:chi2_l1489}, consistent with,
the 9.5 K reported by
Jijina, Adams \& Myers (1999) from observations of NH$_3$.

\subsection{L1517B}

\subsubsection{Best-fit model}

For this cloud we have adopted the same model as described for
L1489-NH3.
The spectra are similar to L1489-NH3 in that they have Gaussian profiles
broadened only slightly above thermal. As in L1489-NH3, the data do
not discriminate between a wide variety of  models with 
different weak velocity fields. 
The best fit model is shown in Figures
\ref{fig:L1517B_Model_777} and \ref{fig:L1517B_Model_777_detail}, 
the model parameters are summarized in Table 3,
and the $\chi^2$ distributions
are given in figures \ref{fig:chi2_l1517b} and  \ref{fig:chi2_surface_l1517b}.  
There are correlations between the abundance and the temperature and core
size similar to those seen in the analysis of L1489-NH3.

\subsubsection{Comparison with other results}

Tafalla \etal\ (2002) included L1517B in their study, finding a central
density of $\sim 2 \times 10^{5}$ cm$^{-3}$ from dust emission, and 
a \diaz\ abundance of $\sim 10^{-10}$ from \diaz\ $J$=1--0 observations. 
While this gas density derived from the dust emission assumes the lower
dust mass opacity of $\kappa_{1.3{\rm mm}} = 0.005$ cm$^2$ g$^{-1}$ and
is consistent with the core gas density that we derive from \diaz\ emission,
the gas {\it column} density derived from the dust is almost
an order of magnitude higher than the gas
{\it column} density derived from the molecular line observations. This is
because the radius of the high density core derived from the dust
emission (Tafalla \etal\ 2004) is a factor of 8 larger than the 
core radius that we derive from
our \diaz\ observations. Thus the molecular line observations again indicate
a higher dust mass opacity closer to 
$\kappa_{1.3{\rm mm}} = 0.04$ cm$^2$ g$^{-1}$. 
Because we have modeled
spectra at only 3 positions in L1517B that are along a strip just off
the center of the cloud,  our estimate of the size of
the core is not well constrained by our data.

\section{Discussion}

In two of the cores in this study, L1489-NH3 and the well studied
core L1544, we find that the density of
molecular hydrogen as determined by a gas tracer (N$_2$H$^{+}$) is about
an order of magnitude below that previously derived from dust observations
if the dust mass opacity is assumed to be 
$\kappa_{1.3{\rm mm}} = 0.005$ cm$^2$ g$^{-1}$. In the third cloud
in our study, L1517B, the estimates of the gas {\it column} density
based on dust and molecular emission are similarly discrepant.
This discrepancy has also been found in other studies.
Chini \etal\ (1993) and Kruegel \& Chini (1994)
computed the mass of the HH24 MMS core via dust (1.3mm emission) and gas
(C$^{18}$O) and found a 2 order of magnitude discrepancy (with the
core mass derived from C$^{18}$O below that estimated from dust).
They argue that freeze-out of C$^{18}$O could not account
for all of this difference. 

Different models of the structure of interstellar dust yield
different estimates of the dust mass opacity. The opacity
of $\kappa_{1.3{\rm mm}} = 0.005$ cm$^2$ g$^{-1}$ is based on
a model of grains in which most of the opacity in the millimeter
wavelengths is due to
silicates (Pollack \etal\ 1994). 
This model might be more
appropriate for dust in a warmer environment with radiative
energy heating the dust.
Kruegel and Siebenmorgen (1994) considered a dust model for colder
environments that
includes the effect of coagulation of dust grains into fluffy
aggregates and the formation of ice mantles around the dust.
They calculate a 
a higher dust mass opacity 
$\kappa_{1.3{\rm mm}} = 0.02$ cm$^2$ g$^{-1}$, that would yield
gas densities from dust more consistent with the molecular line
observations. Ossenkopf and Henning (1994) derived a higher
opacity 
$\kappa_{1.3{\rm mm}} = 1.0$ cm$^2$ g$^{-1}$
for dust composed of fluffy structures and ice mantles.

Our observations suggest that the dust mass opacity in
the starless cores may be close to 
$\kappa_{1.3{\rm mm}} = 0.04$ cm$^2$ g$^{-1}$ 
consistent with grain models 
with enhanced opacities due to fluffy aggregates
and ice mantles that might be found
in the extremely cold, dark interiors of the starless cores.

One way to understand the constraint against higher gas densities
imposed by the molecular line observations is to examine the simple
``critical density'' argument for radiative excitation. Similar to
our
discussion of correlated parameters in section 2.1, this argument is 
intended to be illustrative of the analysis performed by the numerical
radiative transfer modeling rather than an argument in itself. All
the quantitative estimates derived in this study are made 
using the full numerical code and not by order of magnitude arguments.

The observed ratio of the (3--2) and
(1--0) \diaz\ lines suggests a 
gas density sufficiently low that the lines are not
thermalized, and therefore the density must be below
the ``critical density'' for collisional excitation.
For example, the observed corrected antenna temperatures 
of the \diaz(1--0) and \diaz(3--2)
main hyperfine lines in the center position of L1544 are  4.1 K and 1.0 K
respectively. 
Using the brightnesses of the satellite hyperfine lines
in each of the two transitions implies optical depths of roughly 1.7 for 
the (1--0) main hyperfine and 2.7 for the (3--2) main hyperfine.
If both the (1--0) and (3--2) lines were thermalized at a temperature
of 8.5 K, the brightnesses should be about 4.3 and 3.2 K, the latter
significantly higher than the observed (3--2) brightness of 1 K.

The weakness of the (3--2) line implies a low excitation temperature, 
significantly below 8.5 K. 
However, the gas temperature cannot be much lower than 8.5 K
because if it were, the observed 4.1 K brightness of the (1--0) line
would require superthermal excitation. 
The low (3--2) excitation temperature then implies that the density must be low
enough that radiative de-excitation of the (3--2) line is significant
compared to collisional de-excitation. 
The density must be such that 
the collision rate, $n({\rm H}_2)q_{32}$,
is less than $A_{32}$, but $n({\rm H}_2)q_{10}$ is greater than
$A_{10}$. 
The Einstein A coefficients for the (1--0) and (3--2) lines are
$A_{10}=3.62\times 10^{-5}$ s$^{-1}$ and $A_{32} = 1.26\times 10^{-3}$ s$^{-1}$.
 The collisional rate coefficients for \diaz\ should be approximately
the same as those of the similar molecular ion HCO$^+$, that at a gas
temperature of 10 K are, $q_{10} = 2.6 \times 10^{-10}$ cm$^{3}$
s$^{-1}$, and $q_{32} = 4.5 \times 10^{-10}$ cm$^{3}$ s$^{-1}$
(Flower 1999, Table 1).  
Thus to an order of magnitude, 
the gas density in the starless cores should be
between $10^5$ and 
$10^6$ cm$^{-3}$ within most of the
volume of the cloud.

Possibilities other than a higher dust mass opacity that would 
reconcile the difference
between the dust and molecular line estimates cannot be ruled
out. 
The collision rates estimated for N$_2$H$^+$ with
H$_2$ that may be lower than
adopted in our calculations, or alternatively, the \diaz\ 
may be depleted in the centers of the cores.
Our models preclude a 
temperature gradient 
to account for this difference  since a temperature gradient will
only enhance the discrepancy (Evans \etal\ 2001).

In our calculations we have adopted the collision rates 
for HCO$^{+}$ for use with \diaz\ .
(Flower 1999).
Given the similarity in molecular structure HCO$^+$ 
is believed to be a reasonable substitute for \diaz\ (Monteiro 1984).
If the collision process is less efficient
than theoretical estimates, then higher densities would
be required to match the observed line ratios.   
Experimental measurements of the HCO$^+$--H$_2$
collison rates have been perfomed by Oesterling, De Lucia,
\& Herbst (2001), who noted that their
experimental values tended to lie below the theoretical
numbers, but by less than a factor of 2.  However, the experiment included both 
ortho and para forms of H$_2$ so a direct comparison with 
theory, which does not include ortho-H$_2$ as a collision partner, 
was not feasible. 

Current theory suggests that although molecular
ions do not freeze directly onto grain surfaces, depletion of ionic
species from the
gas phase may occur if
the parent molecule (N$_2$ for \diaz) is frozen onto the surfaces of
the dust grains 
(Bergin \etal\ 2002).  If N$_2$H$^+$ were absent in the
densest regions of these cores then the densities could be 
arbitrarily high without any affect on the observed \diaz\ emission.
In our analysis we have examined models where the N$_2$H$^+$ abundance
was allowed to decrease in the central regions. Some of these models
were able to achieve reasonable fits to the data,
but we did not find solutions that improved the fit.  Therefore
we have considered for discussion only the simpler models 
with constant abundance,
but our analysis does not rule out a variable abundance of \diaz .

\section{Conclusions}

1) We have developed a method of analysis to derive the gas
temperature, density, velocity, and molecular abundance from
radio frequency spectral line data by means of 
a radiative transfer model and a heuristic search algorithm
based on simulated annealing.
The radiative transfer model includes a new technique to incorporate 
the hyperfine components  
into the radiative transfer that 
allows acceleration of the $\Lambda$-iteration
with overlapping hyperfine lines.
This technique provides an approximate treatment for
molecules whose collision rates between individual 
hyperfine levels are not known. We also demonstrate
a means to test the accuracy of a radiative transfer code
against a simple semi-analytic calculation.

2) We show how the results of the search for a best fit model
can be used to estimate the uncertainties in the derived
physical data, the temperature, density, velocity, and abundance.
We find that with good enough observational data, that is to say
sufficiently high signal to noise
and angular and spectral resolution, the uncertainties in the
derived results are
not strongly correlated. Rather temperature, density, and abundance
can be estimated independently.

3) As an example, we apply our method of analysis to new observations of 
N$_2$H$^+$ towards
a sample of starless cores, L1544, L1489-NH3, and L1517B.

4) The observed line ratios require that the gas in the cores be
sufficiently rarefied that the \diaz\ lines are not thermalized.
In particular, the H$_2$ density within the cores must be less than 
$\sim 3 \times 10^6$ cm$^{-3}$. Thus the density gradient cannot increase
inward to small radii as a power law with any significant exponent.
We find that models with  constant density at small radii are
consistent with the data.

5) 
The densities derived from the molecular line emission
are lower
than those derived from dust emission if the dust mass opacity
is $\kappa_{1.3{\rm mm}}= 0.005$ cm$^2$ g$^{-1}$, a value suitable
for silicate grains.  A dust mass
opacity of $\kappa_{1.3{\rm mm}}= 0.04$ cm$^2$ g$^{-1}$, suggesting
grain models as fluffy aggregates with ice mantles, yields 
gas densities from the dust observations
that are consistent with the molecular line observations.

6) With the limited data, we are not able to determine whether the
temperature in the starless cores decreases inward or is constant. It is
possible to construct equally good models with constant as well as with
lower central temperatures. Observations at higher angular resolution 
or with additional spectral lines would be required to discriminate
between these models.

7)  The observed velocity fields in L1489-NH3 and L1517B
are consistent with either gravitational
free fall or a static field with microturbulence. 
The \diaz(1--0) spectrum from the central position of L1544 that
shows strong asymmetry that cannot be fit by turbulence alone implying some
inward motion. The best agreement with the data is provided by 
infall that accelerates inward, but the maximum velocity at the
cloud center can be no larger than a few times the thermal velocity.

\acknowledgements

The authors thank Paola Caselli and Mario Tafalla for the use of
their data, and Luca Dore for the calculations of the
\diaz\ hyperfine structure.

\appendix

\section{Comparison Solutions for Spherical Geometry}

For purposes of testing three-dimensional numerical radiative transfer
codes, it is useful to have accurate analytic or semi-analytic
solutions for comparison.  Such solutions can easily give high
accuracy for comparison with numerical codes.

In this appendix we compare our numerical code with the ``standard''
two-level atom problem in a isothermal, constant density sphere.  This
spherical problem can be reduced to an equivalent plane-parallel
problem by a method that originated in the neutron transport
literature (see, e.g., Davison 1957, p.\ 96).  This method was applied
to line transfer in a two-level atom by Sobolev (1962) and 
Cuperman, Engelmann \& Oxenius
(1963, 1964).  The latter authors solved the resulting integral
equation using discrete ordinates in angle and frequency.  Here we use
the kernel approximation method as given by Kunasz \& Hummer (1974).

For plane-parallel media the standard two level atom problem (see, e.g.,
Avrett \& Hummer 1965)
can be formulated in terms of an integral equation with kernel
function
\be
    K(\tau) = {1 \over 2}\int_{-\infty}^{\infty} \phi_x^2 E_1(\phi_x\tau)\,dx
   = {1 \over 2}\int_{-\infty}^{\infty} dx\, \phi_x^2 
             \int_{0}^{1} {d\mu \over \mu} e^{-\phi_x\tau/\mu},  \e{5}
\ee
which is normalized to unity, 
\be
          \int_{-\infty}^{\infty} K(|\tau|)\,d\tau =1.  \e{6}
\ee
The kernel approximation method in plane-parallel geometry 
(Avrett \& Hummer 1965)
involves replacing the
kernel function $K(\tau)$ by an approximation as a sum of
$N$ exponentials:
\be
        K(\tau) = \sum_{i=1}^N a_i e^{-b_i\tau}.   \e{9}
\ee
The normalization of the kernel, as given in Eq.\ (\eqref{6}), implies
that the coefficients of the kernel approximation should satisfy
\be
               \sum_i {a_i \over b_i} = {1 \over 2} .  \e{10}
\ee    
The constants $a_i$ and $b_i$ can be determined by introducing
quadrature rules over $x$ and $\mu$ in the integral (\eqref{5}).
Alternatively, they can be
determined by fitting exponentials to accurate numerical values of the
kernel, as was done by Avrett \& Hummer (1965).  This is especially useful
when the required frequency bandwidth becomes large, as it does for
Voigt profiles with sufficiently large optical depths.  In such cases
kernel fits should achieve a given accuracy in many fewer terms than the
discrete ordinate method.  Extensive kernel fits are given in
Avrett \& Loeser (1966).

We now consider a isothermal, constant density spherical medium of radius $R$
in which the absorption coefficient is spatially constant.  
The mean optical depth from the center to the surface is denoted by
$T$ (the line center optical
thickness $T_0$ is a factor of $1/\sqrt{\pi}$ times this).
We employ the usual notations $B$ 
for the Planck function in the line and $\epsilon$ for the ratio
of $C_{21}/(A_{21}+C_{21})$, where $C_{21}$ is the collisional rate
coeffficient and $A_{12}$ is the Einstein $A$-coefficient for the
upper to lower levels.  

As shown by Kunasz \& Hummer (1974), 
the solution for this standard two-level problem in the sphere
can be written
\be
   S(\tau) =B \left[ 1 - \sum_{j=1}^N L_j {2\sinh(k_j\tau) \over \tau}\right],
           \e{11}
\ee
The $L_j$ and $k_j$ are constants determined by the following equations,
\bea
          1 &=& (1-\epsilon)\sum_i {2a_ib_i \over b_i^2-k_j^2}  \e{13}\\
   {T\over b_i} +{1 \over b_i^2} &=& \sum_j \left({e^{k_j T} \over b_i-k_j}
         -{e^{-k_j T} \over b_i+k_j} \right) L_j \e{14}
\eea
The first condition (\eqref{13}) is the {\em characteristic equation} for the
unknown constants $k_j$.  This is equivalent to an $n$th order
polynomial equation in $k_j^2$, which determines $n$ positive values of
$k_j$.  More practically, the values of $k_j^2$ interleave the
values of $b_i^2$, and may be easily found by Newton's method.

The last condition (\eqref{14}) is then a linear system of equations for the
unknown constants $L_j$, which can easily be solved numerically.  The
source function is then given by Eq.\ (\eqref{11}).

In order to check our three-dimensional radiative transfer code we
considered a set of constant property spherical models for a variety
of values of $\tau$ and $\epsilon$.  The solid curves in 
Figure \ref{fig:analytic} are
accurate results for the source functions using the above semianalytic
method.  We then numerically solved the spherical problem on a uniform
cubical grid, which gave solutions that were approximately spherically
symmetric, as expected.  The numerical results are shown as the dotted
curves in Figure \ref{fig:analytic}
are for a radial ray parallel to the grid.  One
sees that the accuracy of the numerical results is quite good except
near the sharp outer boundary of the sphere, where the true solution
varies over a scale too small to be resolved by our discretization.
We expect better accuracy for cases where the density decreases
outwards more smoothly.

\center{
\begin{table*}
\caption{Model Parameters for L1544}
 
\begin{tabular}{lccc}
\tableline\tableline
Parameter &Best Fit            & Minimum &Maximum  \\
\tableline
 
$T$   (K) & 11  & 7  & 14  \\
${\rm log} n_{\rm core}$ (cm$^{-3}$) & $5.7$ & $5.3$  & $6.0$ \\
$r_{\rm core}$ (pc)  &     0.004  &   0.002   & 0.015 \\
$\alpha$         &     -1.5   &     -2.4   &   -1.25\\
${\rm log} X_{{\rm N}_2{\rm H}^+}$     &  -8.9 &  -9.2     &  -8.5 \\
$v(r)$ (km s$^{-1}$)      &   -0.24  &  -0.30  &  -0.15 \\
$\Delta v_{\rm turb}^a$ (km s$^{-1}$)&  0.0  & ...   & ...  \\
 
 \tableline\tableline \\
 \end{tabular}
 
$^a -$ Microturbulent broadening added in quadrature to the thermal
broadening. Assumed to be zero.
\end{table*}
}

\center{
\begin{table*}
\caption{Model Parameters for L1489-NH3 }
 
\begin{tabular}{lccc}
\tableline\tableline
Parameter &Best Fit            & Minimum &Maximum  \\
\tableline
 
$T$   (K) & 14  & 10  & 17  \\
${\rm log} n_{\rm core}$ (cm$^{-3}$) & $4.9$ & $4.5$  & $5.2$ \\
$r_{\rm core}$ (pc)  &     0.015  &   0.006   & 0.020 \\
$\alpha$         &     -1.5   &     -2.0   &   -1.25 \\
${\rm log} X_{{\rm N}_2{\rm H}^+}$     &  -10.0 &  -10.4     &  -9.5 \\
$v(r)^a$ (km s$^{-1}$)      &   0.0  &  ...  &  ... \\
$\Delta v_{\rm turb}^b$ (km s$^{-1}$)&  0.15   & 0.05    & 0.20  \\
 
 \tableline\tableline \\
 \end{tabular}

$^a -$ Infall velocity assumed to be zero. \hfill\break
$^b -$ Microturbulent broadening added in quadrature to the thermal
broadening.
\end{table*}
}

\center{
\begin{table*}
\caption{Model Parameters for L1517B}
 
\begin{tabular}{lccc}
\tableline\tableline
Parameter &Best Fit            & Minimum &Maximum  \\
\tableline
 
$T$   (K) & 9  & 7  & 15  \\
${\rm log} n_{\rm core}$ (cm$^{-3}$) & $5.1$ & $4.8$  & $5.2$ \\
$r_{\rm core}$ (pc)  &     0.022  &   0.006   & 0.030 \\
$\alpha$         &     -1.4   &     -2.0  &   -0.8 \\
${\rm log} X_{{\rm N}_2{\rm H}^+}$     &  -9.8 &  -10.1     &  -8.0 \\
$v(r)^a$ (km s$^{-1}$)      &   0.0  &  ...  &  ... \\
$\Delta v_{\rm turb}^b$ (km s$^{-1}$)&  0.17  & 0.12   & 0.2  \\
 
 \tableline\tableline \\
 \end{tabular}

$^a -$ Infall velocity assumed to be zero. \hfill\break
$^b -$ Microturbulent broadening added in quadrature to the thermal
broadening.
\end{table*}
}

\clearpage

\begin{figure}[t]
\plotone{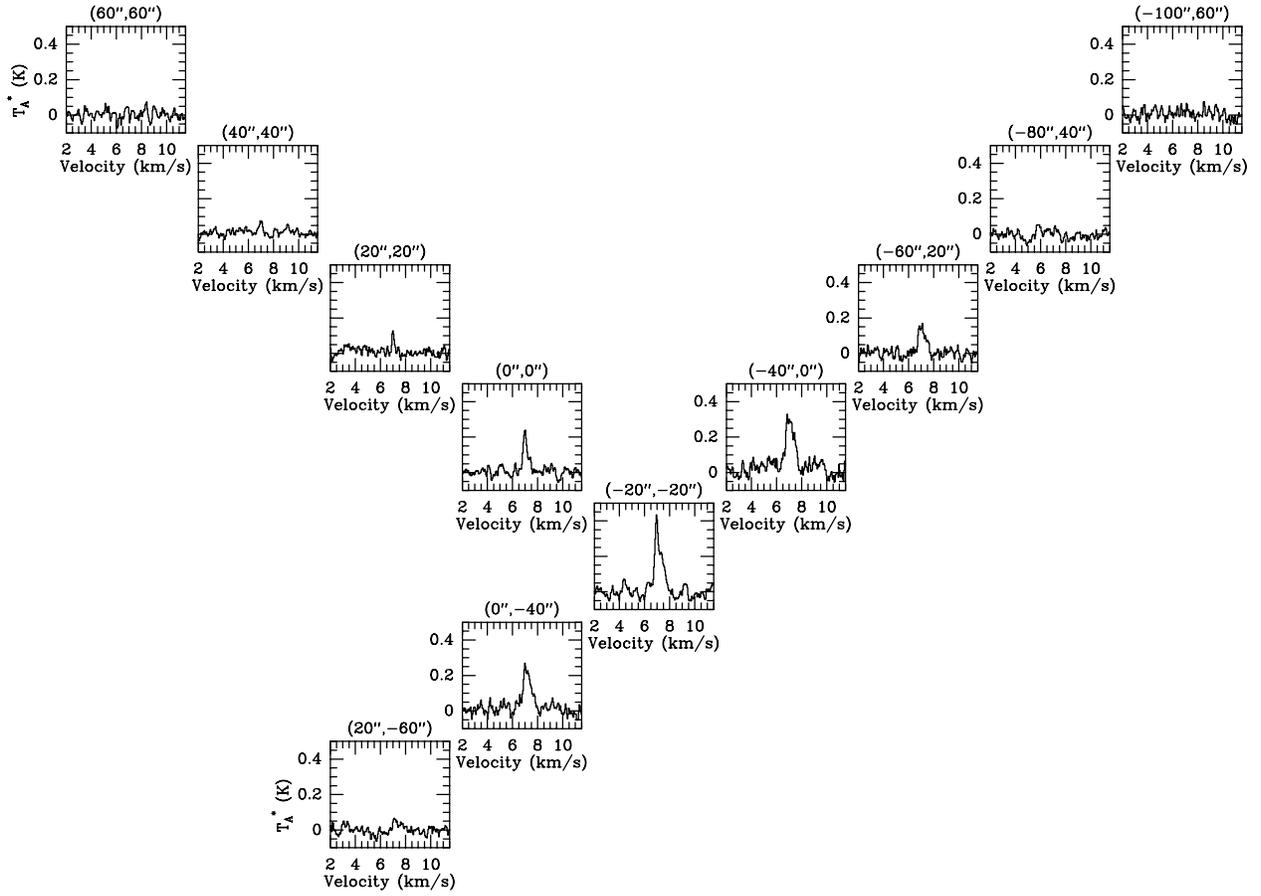}
\caption{\diaz(3--2) spectra at observed positions in L1544.  The spectra are
uncorrected for antenna loss.}
\label{fig:bergin2}   
\end{figure}
\clearpage

\begin{figure}[t]
\epsscale{0.6}
\plotone{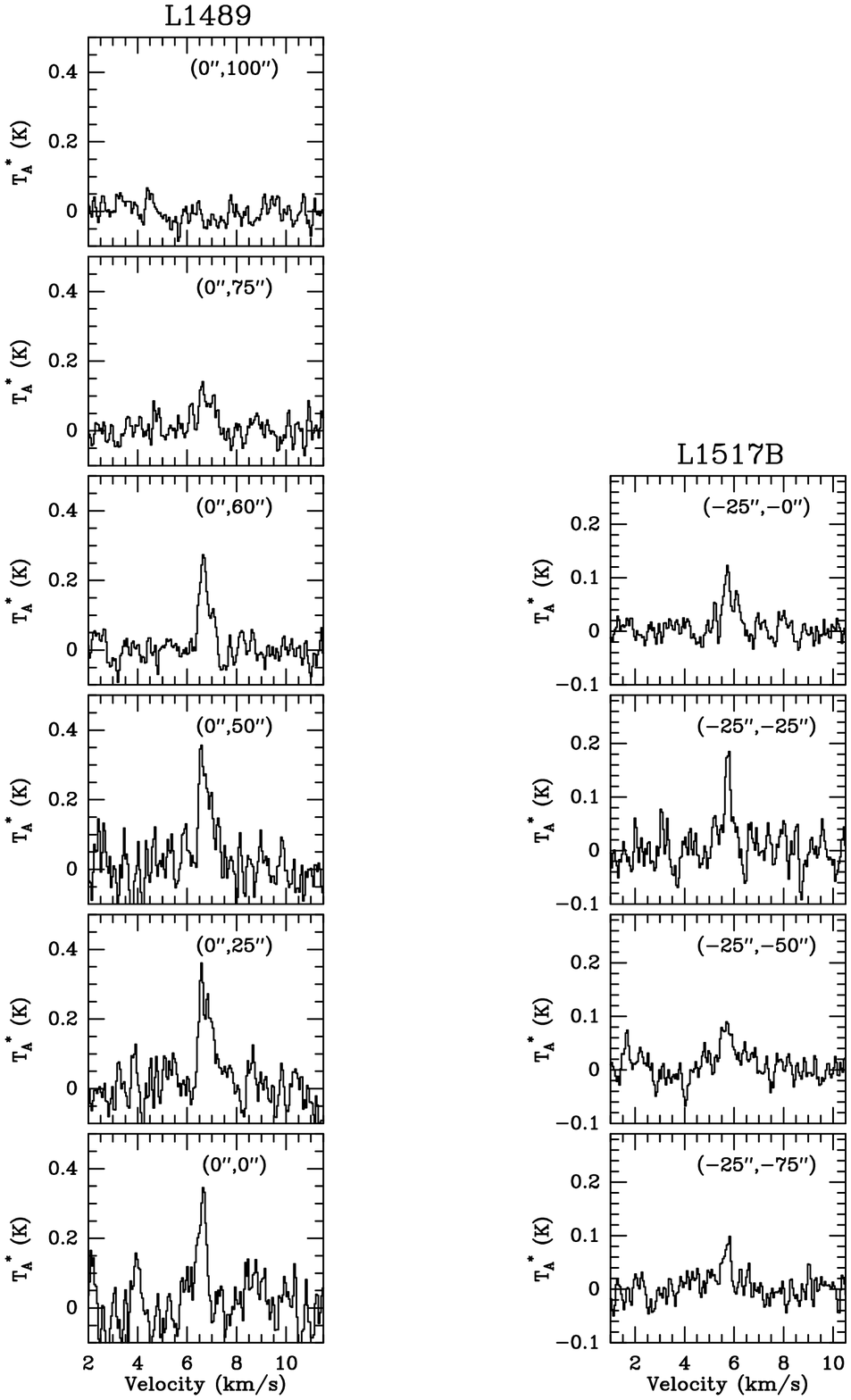}
\caption{\diaz(3--2) spectra at observed positions in L1498 and L1517B.  
The spectra are uncorrected for antenna loss.}
\label{fig:bergin1}  
\end{figure}
\clearpage

\begin{figure} 
\vspace{6.0in}
\includegraphics{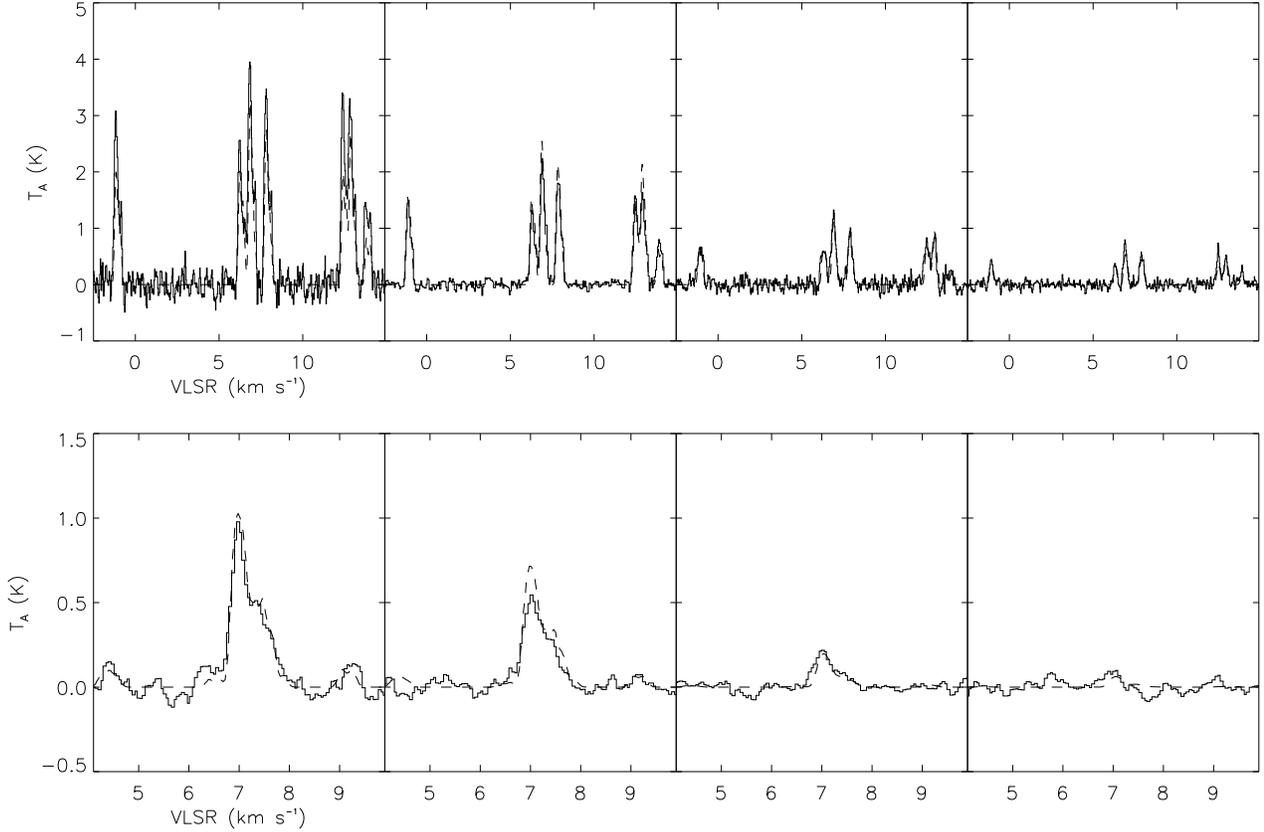}
\caption{Spectra (solid line) and model (dashed line) for
L1544. The top row of spectra are the 3 center hyperfine lines of
\diaz(1--0) at 4 positions across the cloud. The spectrum at the
left is nearest the center of the cloud
(as defined by the dust continuum peak)
while the 3 spectra to the
right are from azimuthal averages of the observed spectra with a
separation between radial positions of 20$^{\prime\prime}$.  
The bottom
row of spectra shows the central hyperfine components of the  \diaz(3--2) line 
from the same positions as the spectra in the first row. The temperature
scale of the data has been corrected for antenna loss and atmospheric
absorption.}
\label{fig:L1544_Model_1213}
\end{figure}
\clearpage

\begin{figure} 
\vspace{6.0in}
\includegraphics{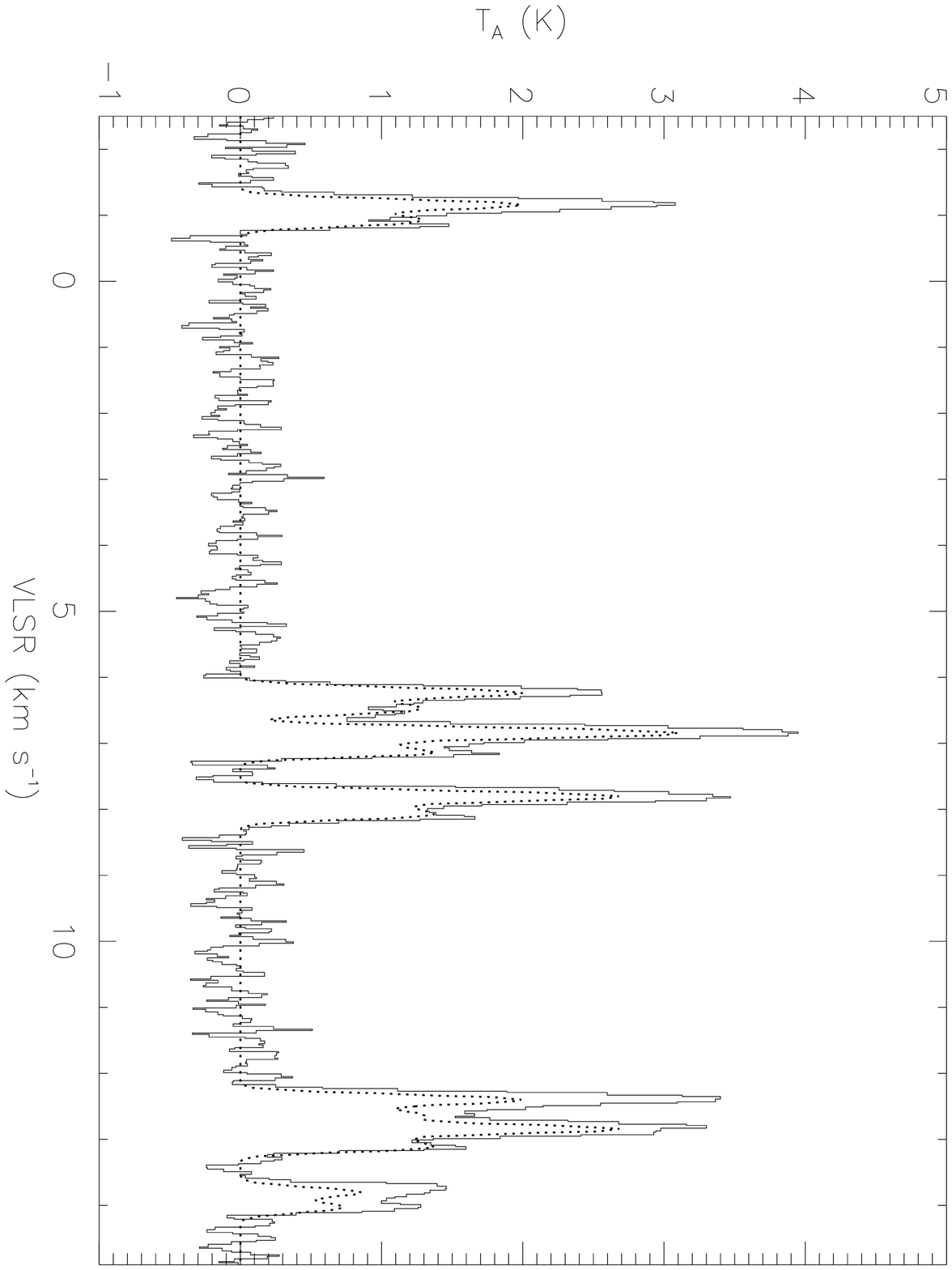}
\caption{First panel of Figure 3(top) (\diaz(1--0), central position)
to show detail. }
\label{fig:L1544_Model_1826_detail}
\end{figure}
\clearpage

\begin{figure} 
\vspace{6.0in}
\includegraphics{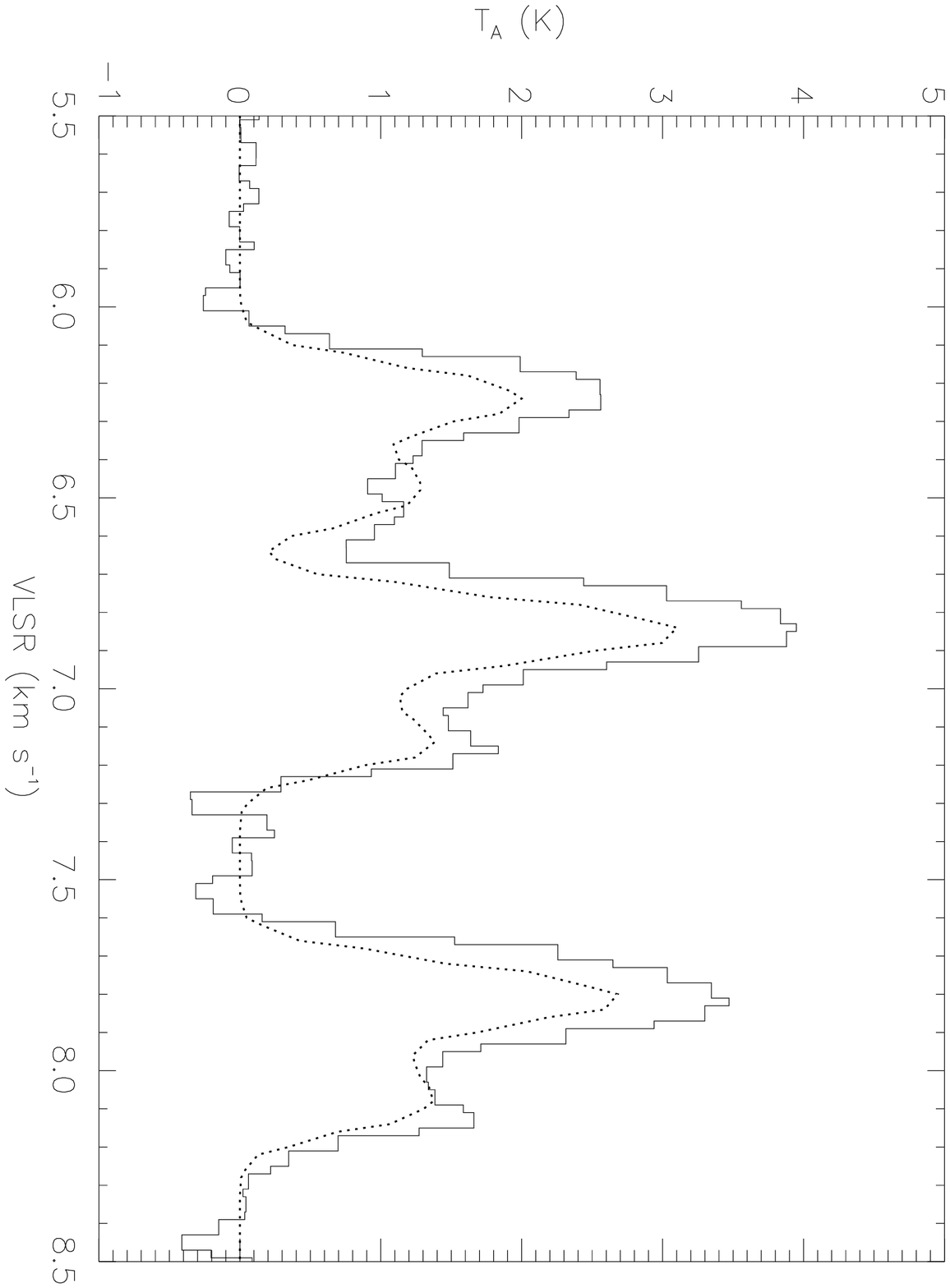}
\caption{First panel of Figure 3(top) (\diaz(1--0), central position)
to show detail. }
\label{fig:L1544_Model_1826_fine_detail}
\end{figure}
\clearpage

\begin{figure} 
\vspace{6.0in}
\includegraphics{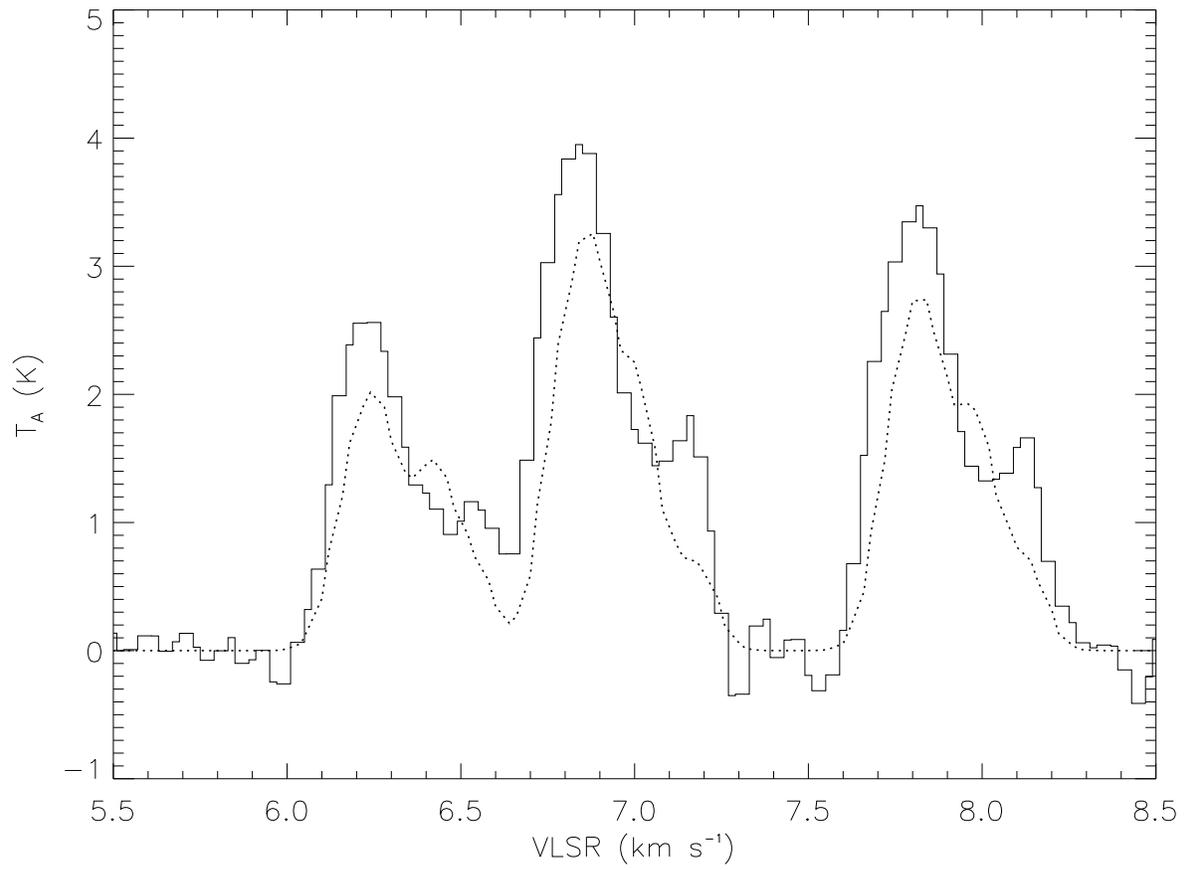}
\caption{Same as figure 5, but from a model with a constant infall velocity
rather than linearly accelerating as in the model of figure 5.
}
\label{fig:L1544_Model_1213_fine_detail}
\end{figure}
\clearpage

\begin{figure}
\plotone{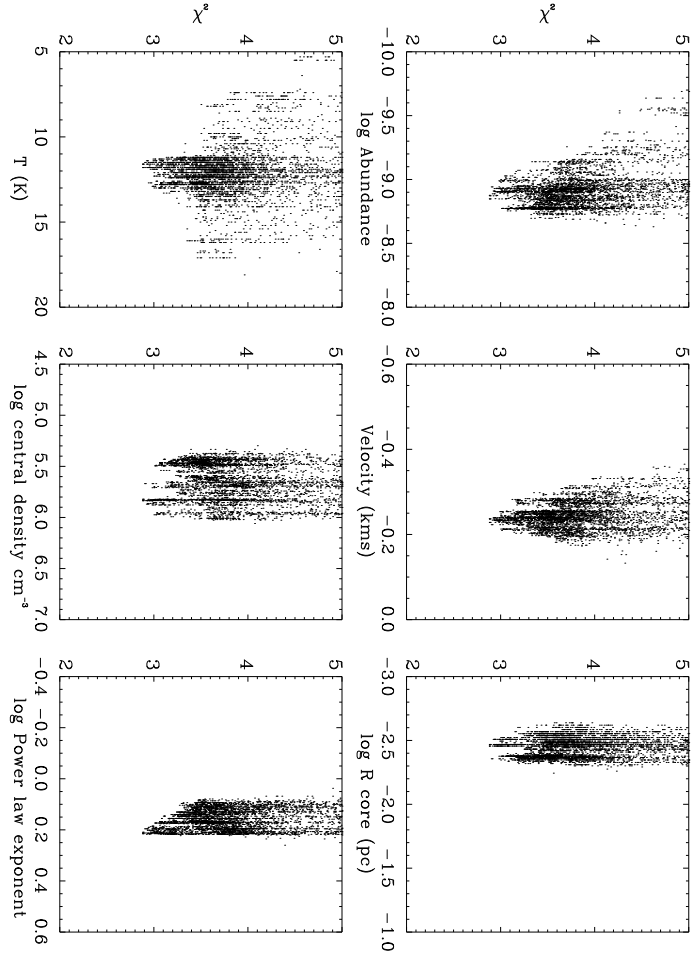}
\caption{Projection of the $\chi^2$ surface on the axes of the model
parameters
for L1544. The lower limit of the points represents an
estimate of the curvature of the $\chi^2$ surface with respect to
variations in each of the model parameters and thus an estimate of
the allowed range of that parameter, or equivalently the
sensitivity of the model to variations in that parameter.}
\label{fig:chi2_l1544}
\end{figure}
\clearpage

\begin{figure}
\vspace{6.0in}
\includegraphics{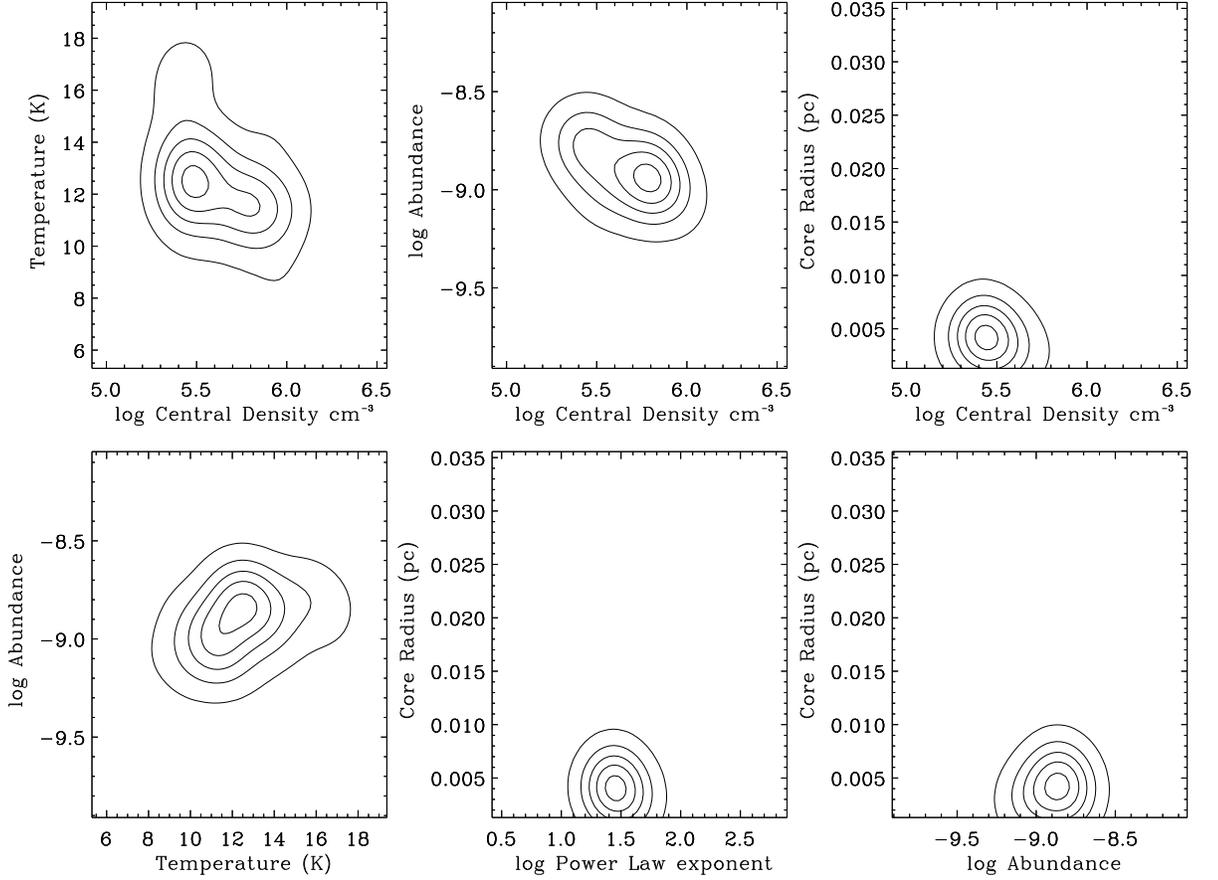}
\caption{The $\chi^2$ surface as a function of some of the
L1544 model parameters taken two at a time. The allowed range for
the parameters is indicated by the curvature of the surface. 
Correlations between parameters would be visible as 
long valleys with an orientation that is not parallel to
either axis. The only significant correlation is between the
central density and the size of the radius of the core.
The ranges shown here have a circular shape indicating that there are
no significant correlations between the model parameters.
The contour values of $\chi^2$ are 3.0 through 4.0 in steps of 0.2.}
\label{fig:chi2_surface_l1544}
\end{figure}
\clearpage

\begin{figure} 
\vspace{6.0in}
\includegraphics{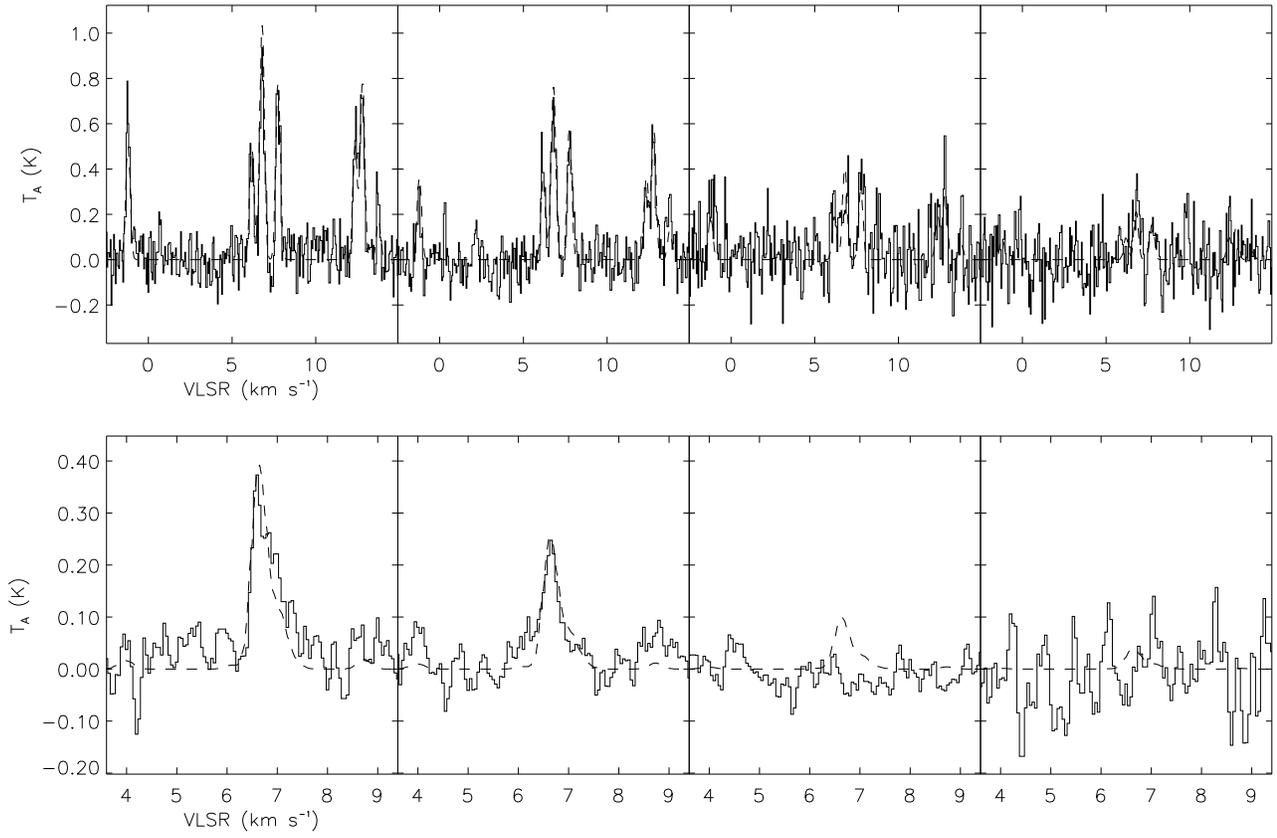}
\caption{Spectra (solid line) and model (dashed line) for
L1489-NH3 in the same format as Figure 3.
}
\label{fig:L1489_Model_955} 
\end{figure}
\clearpage

\begin{figure} 
\vspace{6.0in}
\includegraphics{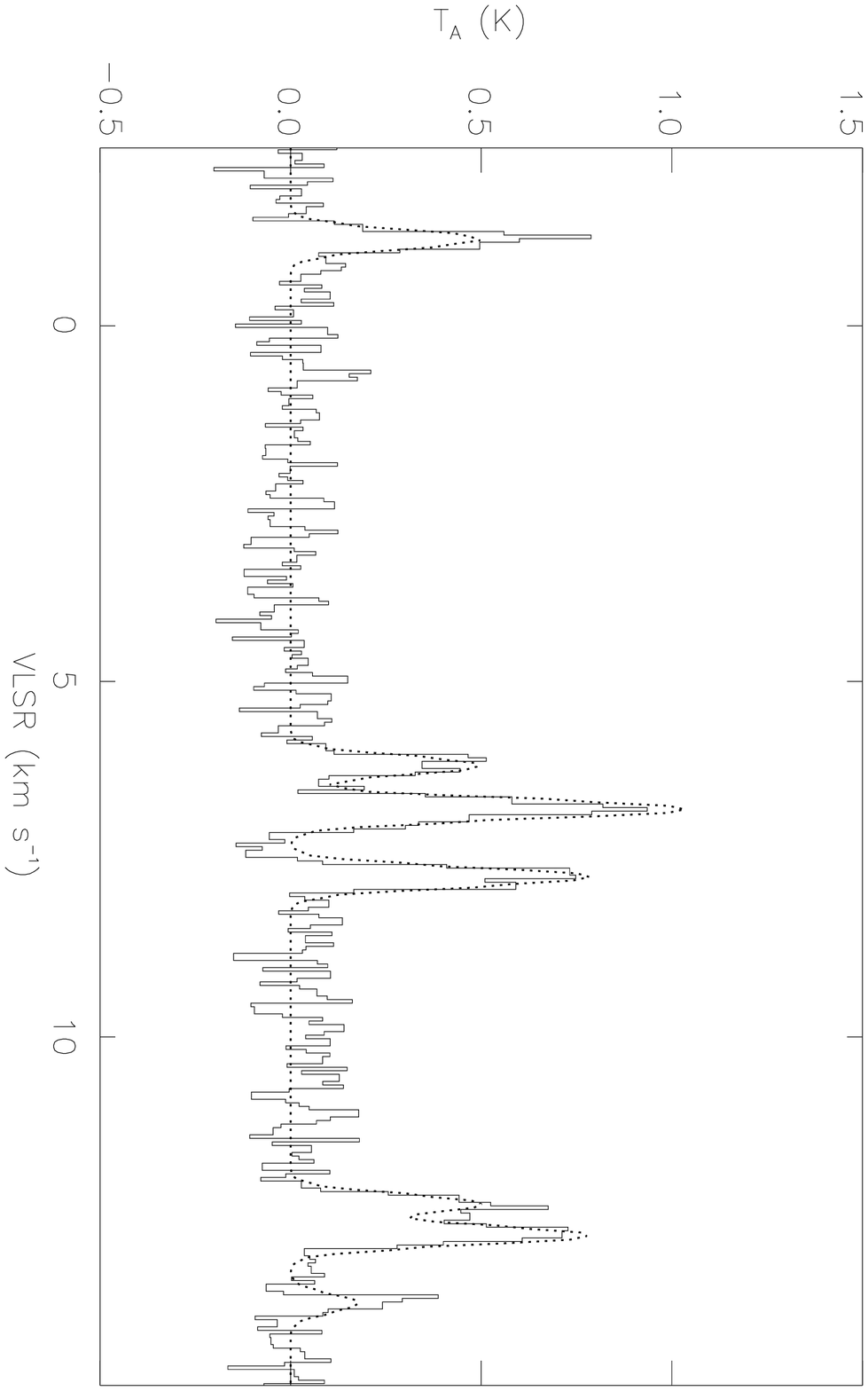}
\caption{First panel of figure 6(top) (\diaz(1--0), central position)
to show detail. }
\label{fig:L1489_Model_955_detail}
\end{figure}
\clearpage

\begin{figure}
\plotone{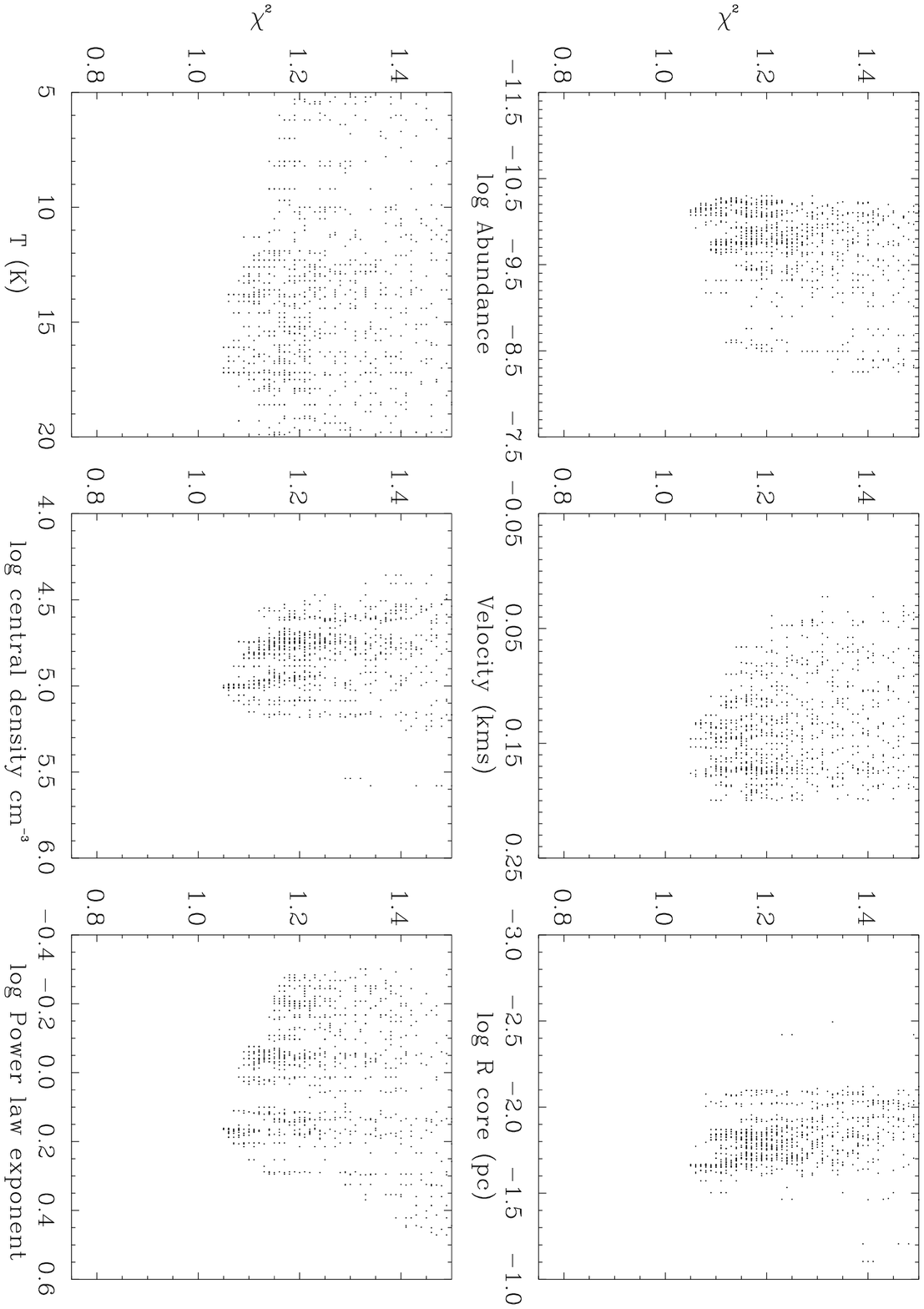}
\caption{Projection of the $\chi^2$ surface on the axes of the model
parameters
for L1489-NH3 in the same format as Figure 5. }
\label{fig:chi2_l1489}  
\end{figure}
\clearpage

\begin{figure}
\vspace{6.0in}
\includegraphics{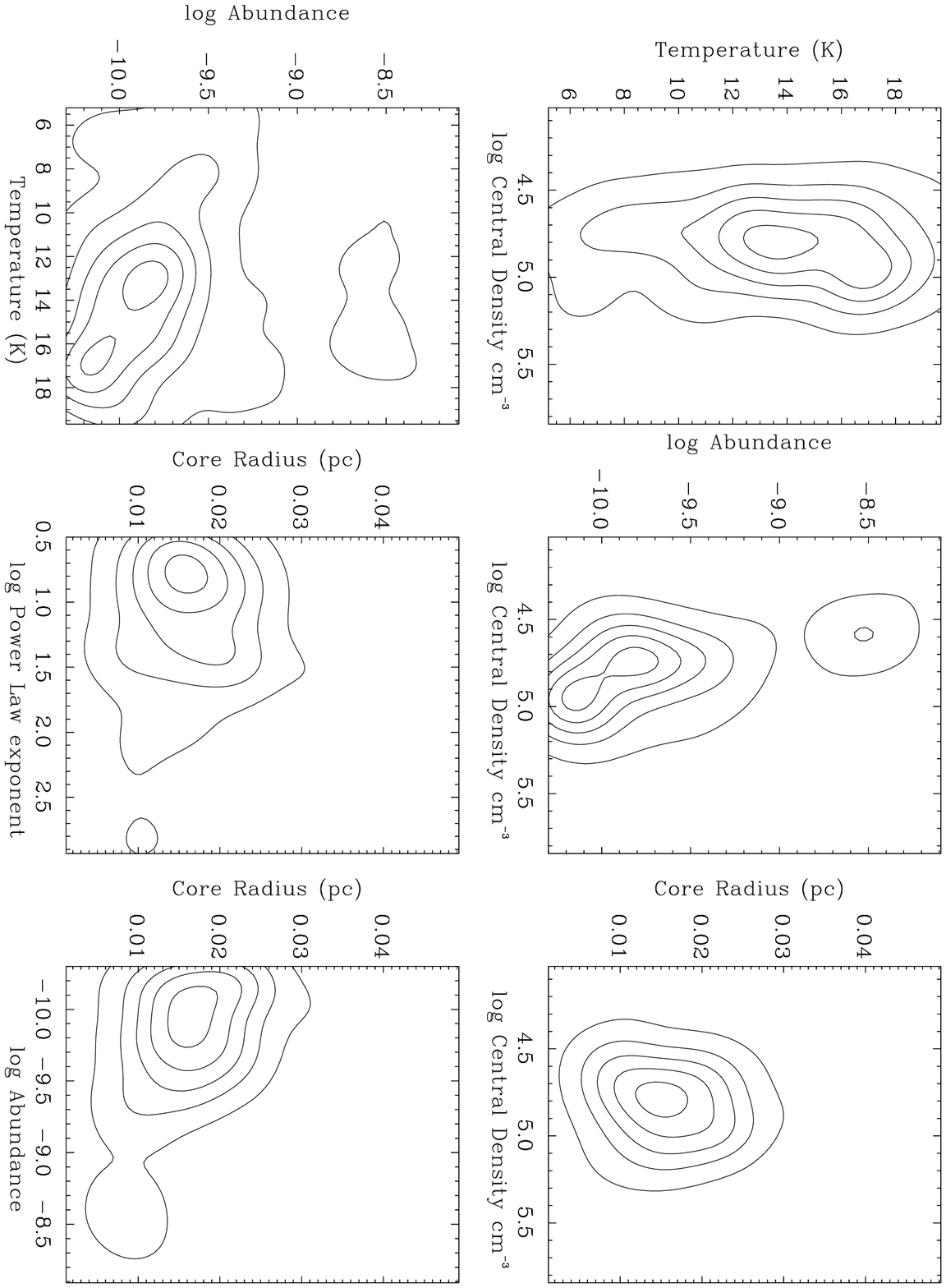}
\caption{
The $\chi^2$ surface for the L1489-NH3 models as a function of some of the
model parameters taken two at a time
in the same format as Figure \ref{fig:chi2_surface_l1544}. 
The contour values of $\chi^2$ are 1.1 through 1.5 in steps of 0.1.}
\label{fig:chi2_surface_l1489}  
\end{figure}
\clearpage

\begin{figure} 
\vspace{6.0in}
\includegraphics{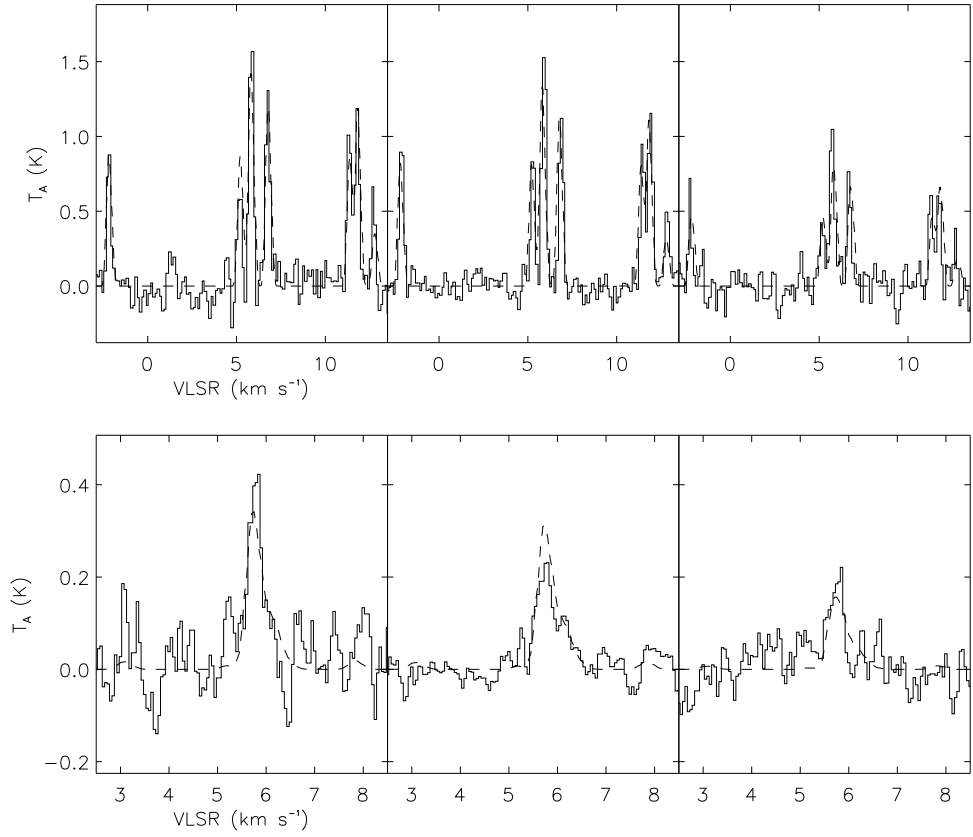}
\caption{Spectra (solid line) and model (dashed line) for
L1517B in the same format as Figure 3.
}
\label{fig:L1517B_Model_777} 
\end{figure}
\clearpage

\begin{figure} 
\vspace{6.0in}
\includegraphics{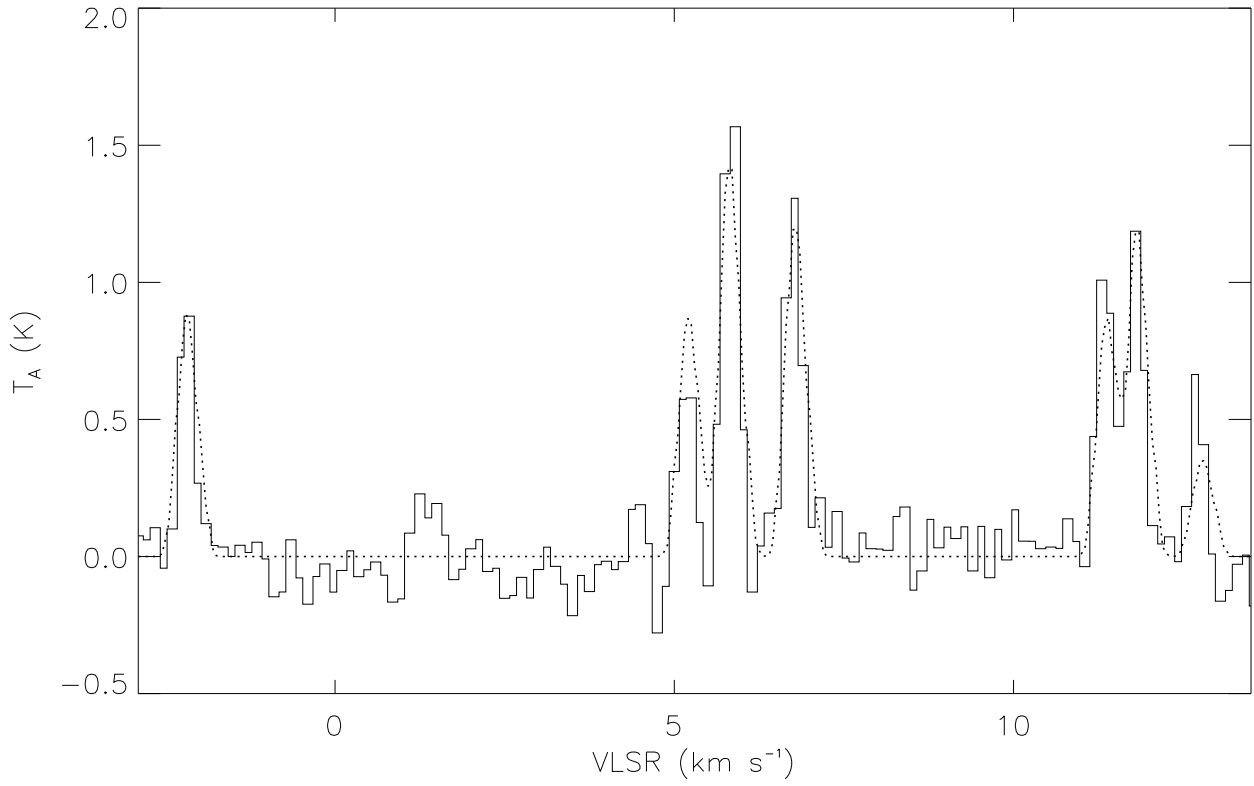}
\caption{First panel of Figure \ref{fig:L1517B_Model_777}
 [\diaz(1--0), central position] to show detail. }
\label{fig:L1517B_Model_777_detail}
\end{figure}
\clearpage

\begin{figure}
\plotone{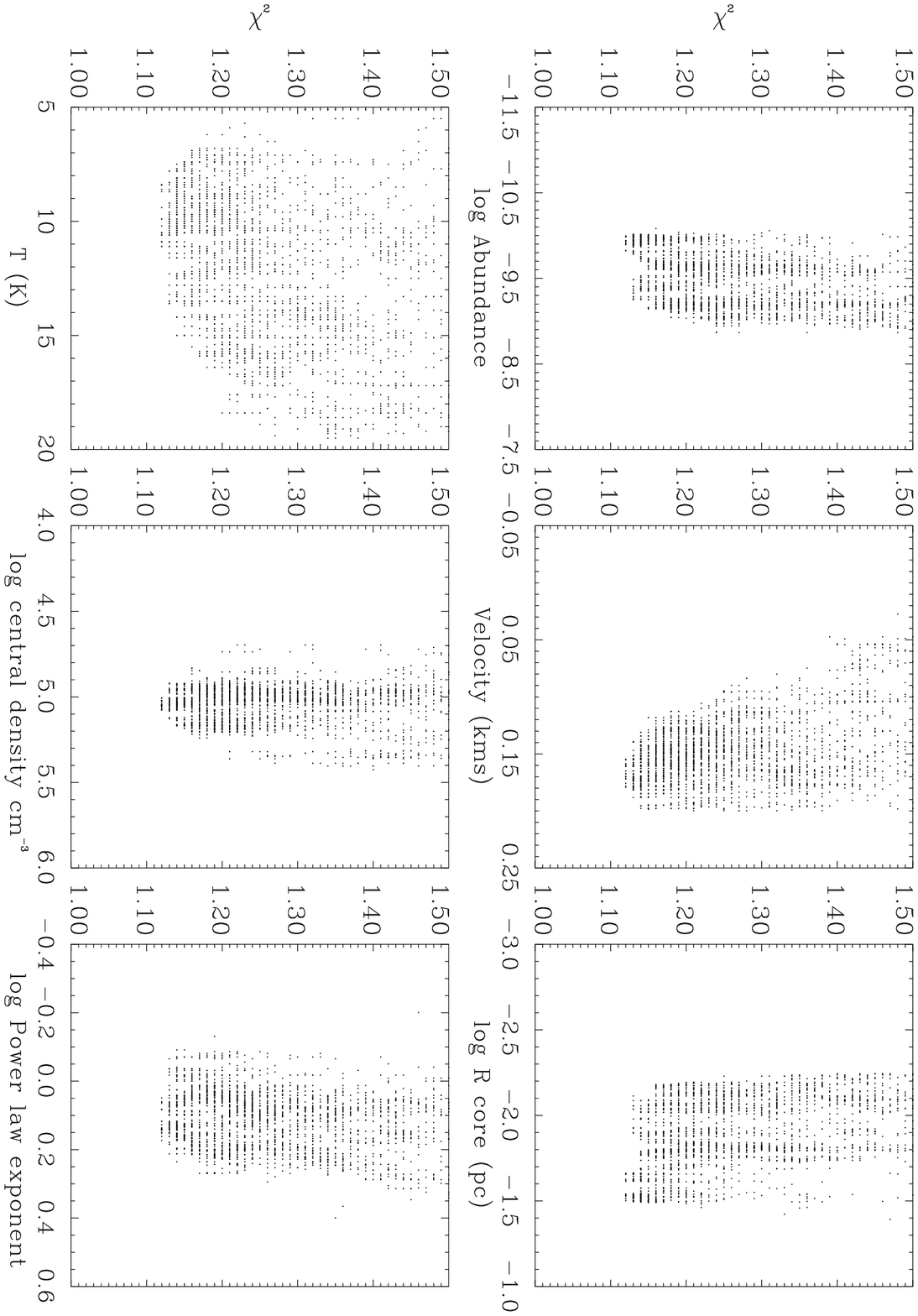}
\caption{Projection of the $\chi^2$ surface on the axes of the model
parameters
for L1517B in the same format as Figure 5.  }
\label{fig:chi2_l1517b}  
\end{figure}
\clearpage

\begin{figure}
\vspace{6.0in}
\includegraphics{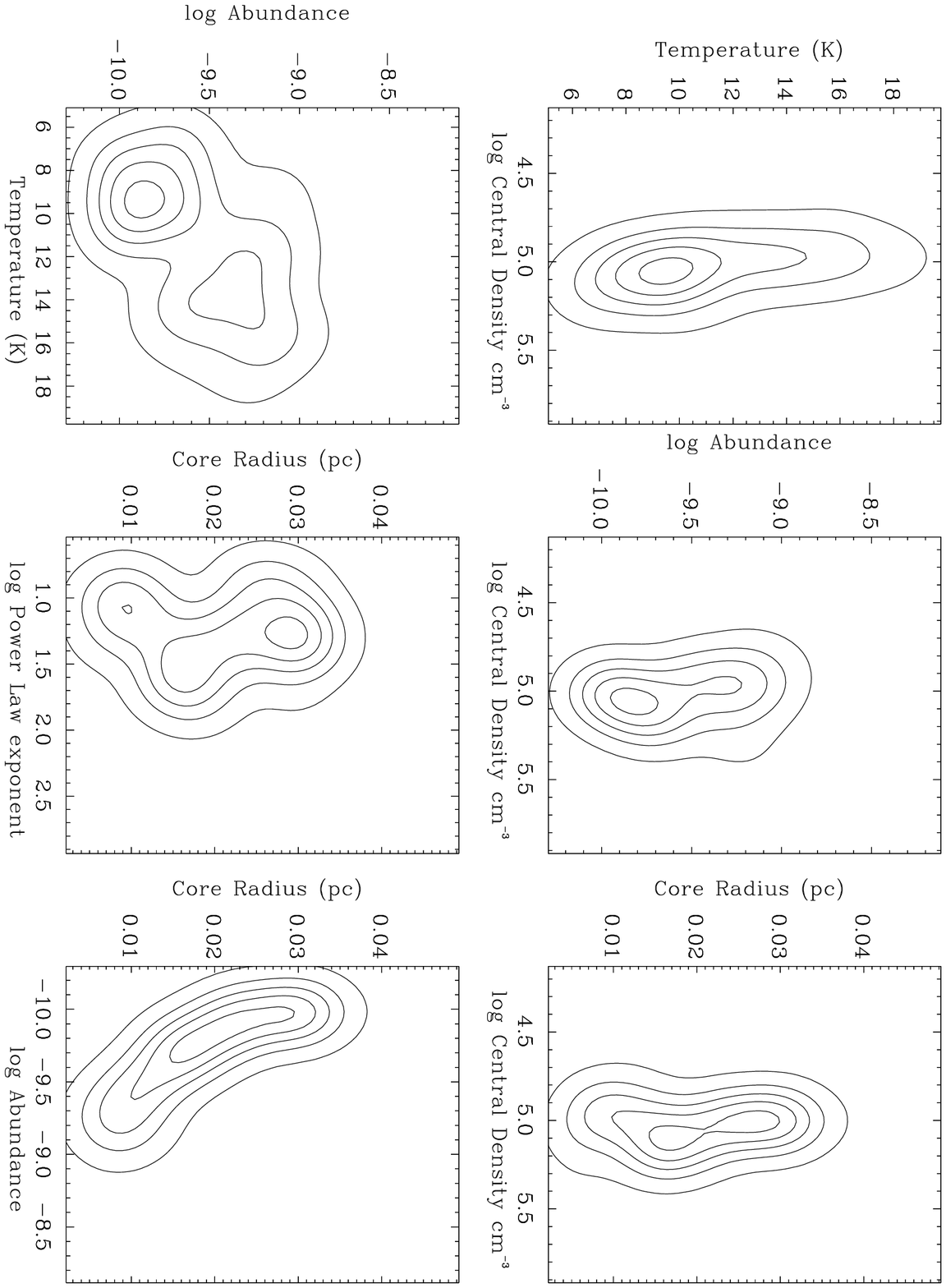}
\caption{
The $\chi^2$ surface for the L1517B models as a function of some of the
model parameters taken two at a time
in the same format as Figure \ref{fig:chi2_surface_l1544}. 
The contour values of $\chi^2$ are 1.1 to 1.5 in steps of 0.2.}
\label{fig:chi2_surface_l1517b}  
\end{figure}
\clearpage

\begin{figure}[t]
\epsscale{0.6}
\plotone{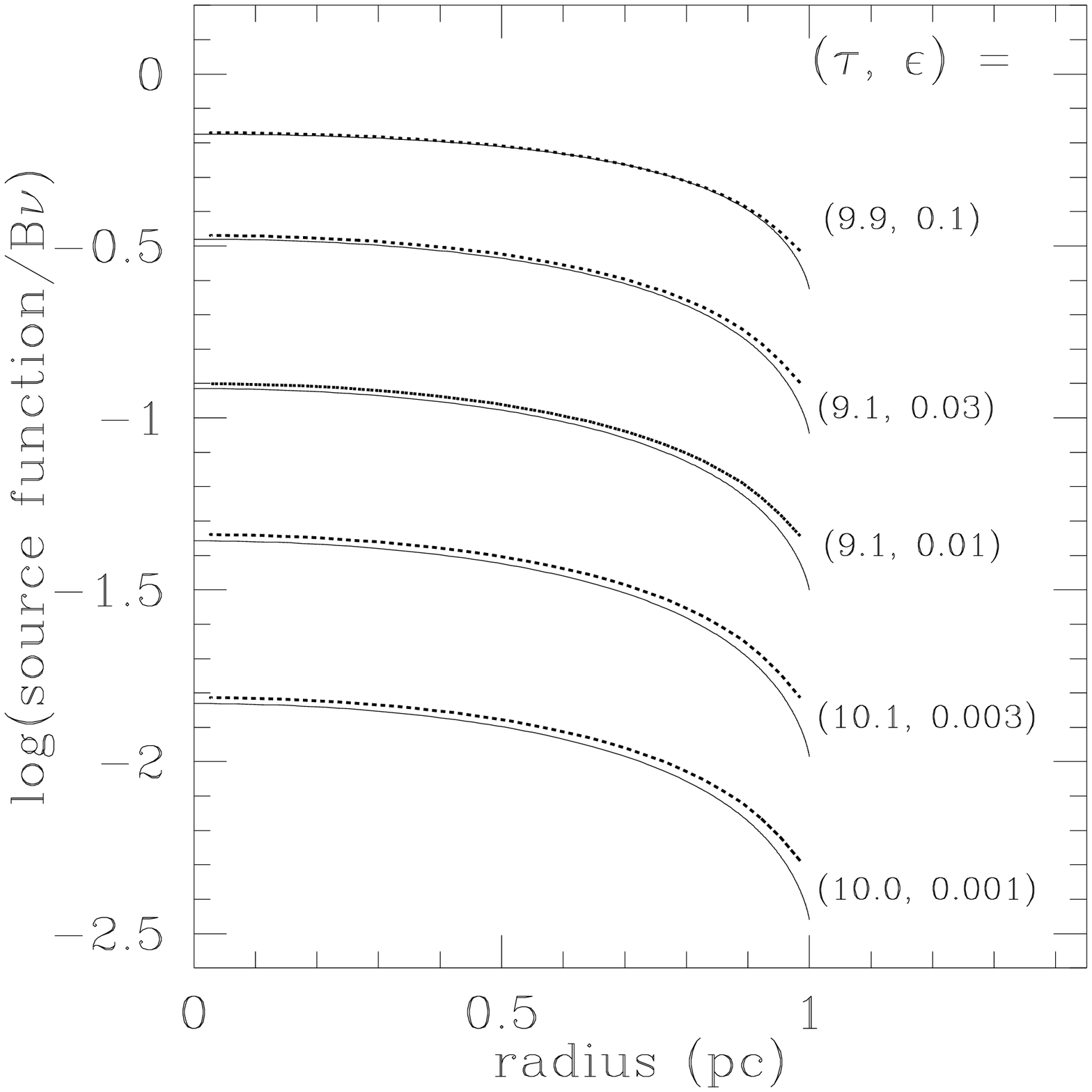}
\plotone{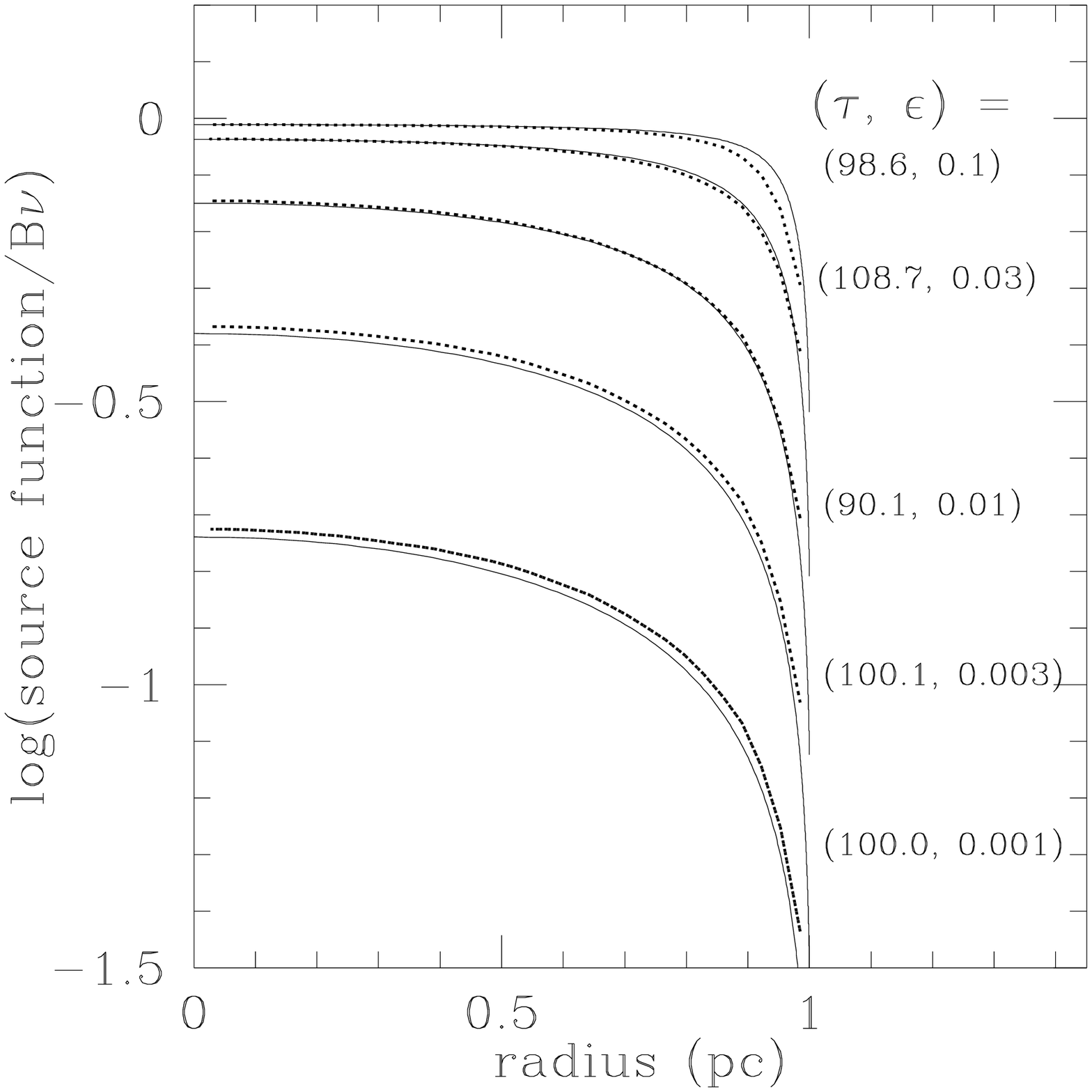}
\caption{Source functions vs.~optical depth for various values of
radial optical depth $\tau$ and $\epsilon$.  Plotted are results using
the accurate semianalytic method (solid) and the numerical code
(dotted).}
\label{fig:analytic}
\end{figure}

\end{document}